\documentclass[a4paper,fleqn]{cas-dc}
\usepackage{lscape}
\usepackage{pdflscape}
\usepackage{float}
\usepackage{stfloats}
\usepackage{algpseudocode}
\usepackage{listings}
\usepackage{xcolor}
\usepackage{caption}
\usepackage{subcaption}
\usepackage{rotating}
\usepackage{algorithm}

\usepackage{xcolor}  
\usepackage{tcolorbox}  
\usepackage[authoryear,longnamesfirst]{natbib}

\begin{document}
\let\WriteBookmarks\relax
\def\floatpagepagefraction{1}
\def\textpagefraction{.001}

\shorttitle{BEACON}


\title [mode = title]{BEACON: A Unified Behavioral-Tactical Framework for Explainable Cybercrime Analysis with Large Language Models}



\author[inst1]{Arush Sachdeva\textsuperscript{1}}

\author[inst1]{Rajendraprasad Saravanan\textsuperscript{1}}

\author[inst1]{Gargi Sarkar\textsuperscript{1}\textsuperscript{*}}

\author[inst1]{Kavita Vemuri}

\author[inst1]{Sandeep Kumar Shukla}

\affiliation[inst1]{
  organization={International Institute of Information Technology Hyderabad},
  addressline={Gachibowli},
  city={Hyderabad},
  postcode={500032},
  country={India}
}

\cortext[0]{\textsuperscript{1}These authors contributed equally to this work.}
\cortext[0]{\textsuperscript{*}Corresponding author. Email: gargi.sarkar@research.iiit.ac.in}






\begin{abstract}
Cybercrime increasingly exploits human cognitive biases in addition to technical vulnerabilities, yet most existing analytical frameworks focus primarily on operational aspects and overlook psychological manipulation. This paper proposes BEACON, a unified dual-dimension framework that integrates behavioral psychology with the tactical lifecycle of cybercrime to enable structured, interpretable, and scalable analysis of cybercrime. We formalize six psychologically grounded manipulation categories derived from Prospect Theory and Cialdini’s principles of persuasion, alongside a fourteen-stage cybercrime tactical lifecycle spanning reconnaissance to final impact. A single large language model is fine-tuned using parameter-efficient learning to perform joint multi-label classification across both psychological and tactical dimensions while simultaneously generating human-interpretable explanations. Experiments conducted on a curated dataset of real-world and synthetically augmented cybercrime narratives demonstrate a 20\% improvement in overall classification accuracy over the base model, along with substantial gains in reasoning quality measured using ROUGE and BERTScore. The proposed system enables automated decomposition of unstructured victim narratives into structured behavioral and operational intelligence, supporting improved cybercrime investigation, case linkage, and proactive scam detection.
\end{abstract}


\maketitle
\section{Introduction}

Cybercrime has emerged as one of the most pervasive and economically destructive consequences of global digitalization. Contemporary online fraud and deception-based crimes now account for unprecedented financial losses worldwide, exceeding trillions of United States dollars (USD) annually \citep{Morgan2016_hackerpocalypse}, while also inflicting severe psychological, social, and reputational harm on victims. Unlike classical cyberattacks targeting systems and networks, modern cybercrime increasingly exploits human vulnerabilities rather than purely technical weaknesses, relying on deception, persuasion, impersonation, emotional coercion, and trust manipulation as primary attack vectors \citep{holt2019human,yao2025psychological, ref_sarkar_behav, ref_sarkar_ttp}.

Existing cybersecurity frameworks, such as the Cyber Kill Chain and the MITRE ATT\&CK framework, provide powerful abstractions for understanding technically sophisticated cyberattacks targeting enterprise systems and critical infrastructure \citep{mitre, capec}. However, these models are fundamentally system-centric: they describe how adversaries compromise digital infrastructure, escalate privileges, and exfiltrate data. In contrast, cybercrime, particularly scams, fraud, impersonation, and extortion, primarily targets individual decision-making processes \citep{louderback2017exploring}, often without exploiting any software vulnerability at all. Consequently, the investigative needs of cybercrime differ substantially from those of traditional cyberattacks. Law enforcement agencies must understand not only how a crime technically unfolds, but also why victims comply with fraudulent demands at different stages of the offense \citep{ref_sarkar_ttp}.

Parallel to these technical models, although not originally formulated for fraud analysis, the theories proposed by \cite{ref_kahneman} provide a foundational explanation of economic decision-making under risk and uncertainty, and have since been widely adopted to characterize the cognitive biases exploited in deceptive and fraudulent interactions. Complementing this perspective, Cialdini’s principles of persuasion provide a social-psychological framework for understanding how influence mechanisms, such as authority, scarcity, and social proof, are systematically leveraged in social engineering and cybercrime \citep{ref_cialdini}. These studies have established that cybercriminals systematically exploit biases such as loss aversion \citep{ridho2024unmasking}, scarcity \citep{de2022poverty, chinecherem2025cybercrime}, authority \citep{fehr2013lure}, reciprocity \citep{falk2006theory}, and emotional dependence \citep{alshammari2023emotional}. 

Recent efforts have introduced Tactics, Techniques, and Procedures (TTP)-based lifecycle models tailored specifically for cybercrime investigation, enabling structured representation of 14 operational stages (see Table \ref{tab:tactics_list} for a full list of tactics) such as reconnaissance, initial contact, credential harvesting, exfiltration, and impact \citep{ref_sarkar_ttp}. Separately, prior studies have examined the role of social influence and cognitive biases in scam victimization \citep{ref_muscanell, ref_ma}. However, no existing automated framework jointly models the operational lifecycle of cybercrime and the underlying psychological manipulation strategies within an explainable, data-driven classification system.

This structural disconnect between technical and psychological perspectives represents a critical gap in current cybercrime research and practice. In real-world investigations, cybercrime does not unfold as a purely technical sequence nor as an isolated psychological manipulation; rather, it manifests as a tightly coupled interaction between adversarial operations and human cognition. Yet, existing analytical frameworks lack the ability to jointly model these interacting dimensions in a single, coherent inferential system. As a result, investigators often rely on manual interpretation of unstructured victim narratives, a process that is slow, inconsistent, and difficult to scale.

\begin{figure*}
    \centering
    \includegraphics[width=0.8\linewidth]{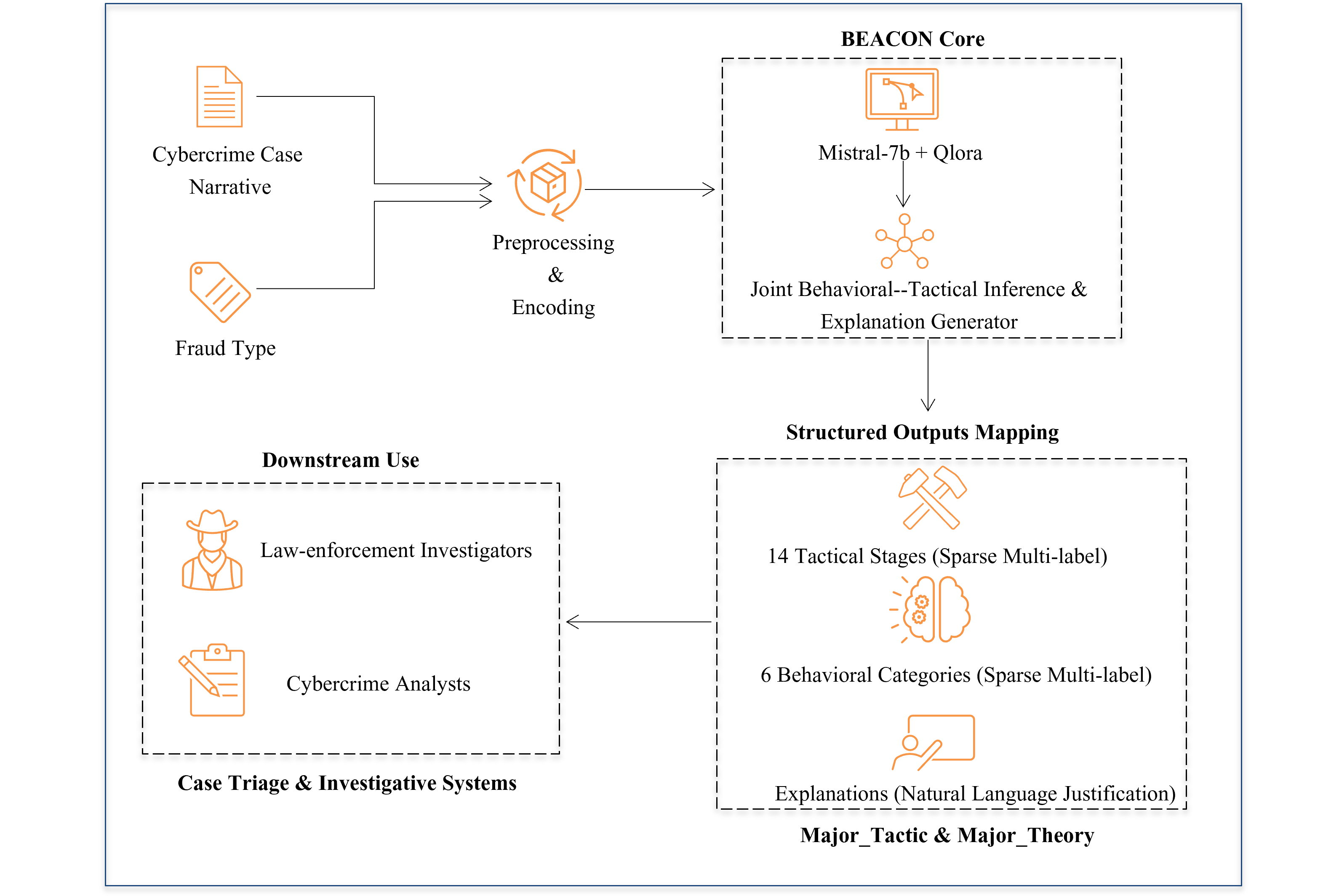}
    \caption{Overview of the BEACON System Architecture}
    \label{beakon}
\end{figure*}

Recent advances in large language models (LLMs) have demonstrated strong capabilities in narrative understanding, reasoning, and structured information extraction from unstructured text. These models offer a promising opportunity to bridge technical and psychological dimensions of cybercrime within a unified analytical framework. However, most existing LLM-based cybercrime studies either focus exclusively on detecting scam content \citep{jiang2024detecting, roy2024chatbots, koide2024chatspamdetector, jamal2024improved, heiding2024devising, koide2024chatphishdetector, lau2012text, bitaab2025scamnet}, classifying high-level fraud types \citep{sarzaeim2024design}, or identifying isolated psychological tactics \citep{meng2024ai, onwuegbuche2025securing, papasavva2025applications, jia2024decision}, without modeling the full operational lifecycle of cybercrime alongside its behavioral drivers. Moreover, few systems provide forensically interpretable explanations alongside predictions, a critical requirement for real-world investigative and legal deployment.

To address these limitations, this paper introduces the first joint behavioral-tactical learning framework for cybercrime analysis that simultaneously infers (i) fine-grained tactical stages of cybercrime execution and (ii) psychologically grounded manipulation strategies directly from unstructured victim narratives. We name the proposed framework \textit{BEACON} (\underline{B}ehavioral--Tactical \underline{E}xplainable \underline{A}nalysis for \underline{C}ybercrime \underline{O}peratio\underline{N}s).
Unlike prior approaches that treat technical progression and psychological manipulation as separate or sequential annotation tasks, the BEACON performs true cross-dimensional, multi-label joint inference within a single end-to-end model. This enables the system to determine not only which operational stage of cybercrime is present, but also which cognitive vulnerabilities are being exploited at that stage, thereby linking adversarial strategy with human decision-making in a inferentially interpretable manner.

To support this formulation, we construct a dual supervised cybercrime dataset in which each incident is simultaneously annotated with fourteen tactical stages of cybercrime execution and six psychologically grounded behavioral manipulation categories, each accompanied by concise explanatory rationales. This enables both predictive classification and explanation-alignment evaluation under a unified supervision scheme. Building on this dataset, we fine-tune an LLM to perform joint multi-label behavioral-tactical classification with explanation generation, producing structured, human-verifiable outputs suitable for investigative workflows.

Through comprehensive experimental evaluation on real-world cybercrime narratives, we demonstrate that domain-adapted joint learning substantially outperforms zero-shot LLM inference across both tactical and psychological dimensions, while also yielding high-quality explanations that closely align with human reasoning. BEACON enables structured decomposition of complex cybercrime cases into operational stages and behavioral levers, facilitating improved case triage, pattern discovery, and investigative consistency. Figure \ref{beakon} illustrates the end-to-end architecture of the proposed BEACON framework for joint behavioral-tactical cybercrime analysis.

\subsection*{Contributions}

This work makes the following principal contributions:

\begin{enumerate}
    \item \textbf{First LLM-driven automation of TTP-based cybercrime lifecycle with joint behavioral inference.}
    Building on this foundation of TTPs for Cybercrime Investigation introduced by \cite{ref_sarkar_ttp}, we present the \emph{BEACON}, the \emph{first LLM-driven automated instantiation of this tactical cybercrime framework}, extended with simultaneous inference of psychologically grounded manipulation strategies. It performs true cross-dimensional, multi-label joint inference from unstructured crime narratives within a single end-to-end model, transforming a previously analyst-driven framework into a scalable, machine-assisted investigative intelligence system.

    \item \textbf{Dual-supervised cybercrime dataset with explanatory annotations.}
    We introduce a dual-supervised cybercrime corpus in which each incident is annotated with (i) 14 operational cybercrime tactics, (ii) 6 psychological manipulation categories, and (iii) human-verifiable explanatory rationales for each label. To the best of our knowledge, this is the first dataset to provide simultaneous tactical--behavioral supervision together with explanation alignment, enabling research on both predictive performance and interpretable cybercrime analysis. Although the underlying cognitive manipulation strategies are domain-general, the linguistic realization and scam narratives in the dataset partially reflect India-specific socio-technical and cultural contexts, including the use of local institutional impersonation (e.g., police, tax authorities), regionally prevalent payment infrastructures (e.g., UPI-based transfers), and Indian English stylistic patterns. These contextual features are preserved to ensure ecological validity for the target deployment setting, while the higher-level tactical and behavioral abstractions remain broadly transferable across geographic regions.

  \item \textbf{Comprehensive evaluation of joint inference and explainability.} 
  We conduct a systematic empirical evaluation of the BEACON on a held-out test set of real-world cybercrime narratives sourced from verified media reports and human-annotated cases, benchmarking a domain-adapted LLM against its zero-shot baseline across all 20 tactical and psychological labels. In addition to standard multi-label classification metrics, we evaluate explanation quality using ROUGE and BERTScore, demonstrating that joint supervision substantially improves both predictive accuracy and forensic reasoning fidelity.

\end{enumerate}

The rest of this paper is organized as follows. Section~\ref{sec:relatedwork} reviews relevant background and related work. Section~\ref{sec:method} describes the dataset and modeling methodology. Section~\ref{sec:experiments} presents the experimental setup and results. Section~\ref{sec:discussion} discusses key findings and limitations. Section~\ref{sec:conclusion} concludes the paper with directions for future research.

\
\section{Background and Related Work}
\label{sec:relatedwork}

\subsection{Cybercrime as a Socio-Technical Phenomenon}

\begin{table*}
\centering
\caption{Tactics in the Cybercrime Lifecycle by \cite{ref_sarkar_ttp}}
\label{tab:tactics_list}
{\fontsize{9}{11}\selectfont
\begin{tabular}{p{3.2cm} p{5.0cm} p{6.0cm}}
\toprule
\textbf{Tactic} & \textbf{Description} & \textbf{Example (Cybercrime Context)} \\
\midrule
Reconnaissance & Collecting personal information about the victim to prepare the scam. & Buying leaked data, scraping social media profiles for names, phone numbers, and family links. \\

Resource Development & Acquiring digital or human resources required to commit cybercrime. & Renting fake call center spaces, purchasing SIM cards and spoofing applications, hiring money mules. \\

Initial Contact & Initiating deceptive communication with the victim. & Romance-scam direct messages, fake technical support calls, fraudulent job-offer SMS messages. \\

Detonation & Triggering the fraudulent act by inducing the victim to take a harmful action. & Convincing the victim to share OTPs, install spyware, or transfer money. \\

Persistence & Maintaining prolonged interaction to maximize exploitation. & Repeated scam calls, sustained emotional manipulation, continued TeamViewer access. \\

Escalation & Transitioning from low-level fraud to higher-stakes criminal activity. & Using cloned fingerprints for AEPS fraud after initial data compromise. \\

Defense Evasion & Obscuring the offender’s identity, location, or digital trace. & Use of VoIP numbers, VPN services, and cryptocurrency for laundering. \\

Credential Harvesting & Extracting sensitive authentication or personal credentials. & Deploying keyloggers, gaining banking passwords through romance-based manipulation. \\

Discovery & Gaining deeper access to victim data for subsequent exploitation. & Accessing contact lists, photo galleries, and messages to extract blackmail material. \\

Pivoting & Leveraging one compromised victim to reach additional victims. & Sending scam SMS messages to contacts stored on an infected device. \\

Collection & Capturing valuable information from victim devices. & Recording calls, accessing WhatsApp chats through spyware. \\

Command and Control & Directly controlling or coercing victim behavior. & Threatening disclosure of private material unless payment is made (sextortion). \\

Exfiltration & Illegally transferring funds or sensitive data out of victim control. & Sending stolen money to mule accounts or data to criminal Telegram groups. \\

Impact & Social, financial, or psychological harm inflicted on the victim. & Monetary loss, identity theft, suicide due to blackmail, defamation through leaked images. \\
\bottomrule
\end{tabular}
}
\end{table*}

Cybercrime differs fundamentally from conventional cyberattacks in that it is predominantly \textit{human-centric rather than system-centric} \citep{holt2019human,  ref_sarkar_ttp, ref_sarkar_behav, david2023overview,hagen2025human, louderback2017exploring, van2018deviating, yao2025psychological}. While classical cyber intrusions exploit software vulnerabilities, misconfigurations, and cryptographic weaknesses, deception-driven cybercrime primarily targets \textit{human cognition, perception, and trust}. Scams, fraud, impersonation, and extortion succeed not only because of flaws in digital infrastructure alone, but also because victims are socially engineered \citep{syafitri2022social, louderback2017exploring, yao2025psychological} into voluntarily executing harmful actions, such as sharing credentials, transferring funds, or installing malware. 

Moreover, the manifestation and effectiveness of such deception-driven crime are shaped by socio-economic, institutional, and cultural contexts. In the Indian setting, for example, strong institutional trust in government authorities, informal credential-sharing practices within trusted social networks, and widespread reliance on centralized digital identity and payment infrastructures (e.g., UPI) frequently influence how cognitive exploitation is operationalized. By contrast, in Western contexts, stricter privacy norms, higher regulatory enforcement, and greater adversarial skepticism toward institutions often shape different scam modalities. These variations highlight that while the underlying cognitive mechanisms of deception are broadly universal, their operational realization is fundamentally context dependent.

Overall, this multifaceted nature of cybercrime as both a \textit{technical process} \citep{ref_sarkar_ttp} and a \textit{psychological manipulation process} \citep{david2023overview,hagen2025human, louderback2017exploring, van2018deviating, coluccia2020online} necessitates analytical frameworks that simultaneously account for (i) the operational progression of the offense, and (ii) the cognitive mechanisms that enable victim compliance. 

\subsection{Technical Lifecycle Modeling of Cybercrime}

The modeling of attacks as sequences of adversarial objectives originated in military doctrine and was formalized for cybersecurity through the \textit{Cyber Kill Chain}, which describes stages from reconnaissance to actions on objectives \citep{Lockheed}. The \textit{MITRE ATT\&CK} framework subsequently extended this concept into a comprehensive matrix of tactics and techniques used in enterprise-level and nation-state intrusions \citep{mitrepaper}. These frameworks are highly effective for system-centric intrusions, where attackers compromise hosts, escalate privileges, and exfiltrate data through technical exploitation. However, they are not designed to model cybercrime targeting individual victims, where the primary attack surface is the human user and the dominant vector is social engineering \citep{ref_sarkar_ttp}.

To address the gap between intrusion-centric models and cybercrime investigation needs, Sarkar and Shukla introduced a \textit{TTP-based cybercrime lifecycle framework} specifically tailored for cybercrime investigation. This model conceptualizes cybercrime as a sequence of a subset of fourteen operational tactics: \textit{Reconnaissance, Resource Development, Initial Contact, Detonation, Persistence, Escalation, Defense Evasion, Credential Harvesting, Discovery, Pivoting, Collection, Command and Control, Exfiltration, and Impact} (see Table \ref{tab:tactics_list}). Each tactic represents a distinct adversarial objective rather than merely a technical action. For example, \textit{Reconnaissance} involves profiling potential victims using leaked databases or social media, whereas \textit{Detonation} corresponds to the cognitive trigger point at which the victim is induced to take a harmful action such as transferring money or disclosing credentials. Importantly, this framework models cybercrime as a progressive socio-technical process rather than a single isolated event.

However, the TTP framework for cybercrime investigation was proposed as a \textit{manual, analyst-driven investigative aid}. It does not provide automated inference, machine learning support, or scalable real-time decision support, thereby limiting its deployment in large-scale cybercrime analysis.

\begin{table*}
\centering
\caption{Taxonomy of Behavioural Theories in Cybercrime}
\label{tab:behavioural_theories}
{\fontsize{9}{11}\selectfont
\setlength{\tabcolsep}{6pt}
\begin{tabular}{p{0.35\linewidth} p{0.55\linewidth}}
\toprule
\textbf{Category} & \textbf{Theoretical Foundation and Mechanism} \\
\midrule

\textbf{Fear and Intimidation} &
Rooted in loss aversion and negative affect under Prospect Theory. Captures threats, coercion, or fear-inducing stimuli, where victims overweight potential losses and comply to avoid perceived negative outcomes. \\

\textbf{Urgency and Scarcity} &
Based on Cialdini’s scarcity principle. Leverages deadlines, expiring opportunities, and high-pressure scenarios, where a perceived reduction in time disrupts rational evaluation and increases impulsive compliance. \\

\textbf{Authority, Social Proof, and Impersonation} &
Derived from Cialdini’s authority and social proof principles. Reflects impersonation of trusted institutions (e.g., banks, police) or fabricated consensus, causing victims to defer to perceived expertise or majority behavior. \\

\textbf{Consistency and Reciprocity} &
Based on Cialdini’s commitment/consistency and reciprocity principles. Exploits the human tendency to honor prior commitments or repay favors, where small initial requests escalate into larger exploitative actions. \\

\textbf{Phantom Riches} &
Grounded in Prospect Theory’s value function. Refers to exaggerated rewards such as fictitious prizes or unusually high investment returns, exploiting the tendency to overweight low-probability but high-gain outcomes. \\

\textbf{Emotional Exploitation} &
Captures the affective dimension central to modern scams (e.g., romance and sextortion fraud). Uses empathy triggers, emotional narratives, and relational bonding to bypass analytic judgment and increase vulnerability. \\

\bottomrule
\end{tabular}
}
\end{table*}

\subsection{Psychological Foundations of Cybercrime}

Cybercrime differs from conventional cyberattacks in that its success is fundamentally rooted in the exploitation of human cognition, emotion, and social behavior rather than purely technical system vulnerabilities \citep{ref_sarkar_ttp, ref_sarkar_behav}. A rich body of behavioral science has therefore been leveraged to explain why individuals become susceptible to deception-driven cybercrime.

From the perspective of the victim, Prospect Theory explains how individuals systematically deviate from rational decision-making under uncertainty by overweighting potential losses and improbable large gains \citep{ref_kahneman}. Cybercriminals exploit loss aversion in threat-based extortion and digital-arrest scams, as well as gain an advantage in high-return investment and lottery frauds. These cognitive distortions impair risk evaluation and enable attackers to bypass analytical reasoning even among technologically literate victims. However, decision-theoretic models such as Prospect Theory primarily operate once an individual has already engaged with a risky proposition. Initial victim engagement is often shaped by dispositional and situational factors such as financial ambition, perceived opportunity, trust, emotional vulnerability, or socio-economic stress, which determine baseline susceptibility prior to formal risk evaluation.

From the perspective of the fraudster, deception is operationalized through systematic influence strategies designed to elicit compliance and suppress critical reasoning. Furthermore, Cialdini’s principles of persuasion, authority, scarcity, reciprocity, commitment and consistency, social proof, and liking, constitute the core psychological arsenal of modern social engineering and cybercrime \citep{ref_cialdini}. These principles are deliberately orchestrated to induce urgency, legitimacy, trust, and obligation, thereby shaping victim behavior toward attacker-favorable actions such as in bank and police impersonation (authority), artificial deadlines and account-closure threats (scarcity), fake refunds and assistance scams (reciprocity), and emotional grooming in romance and sextortion fraud (liking and consistency). These manipulation strategies are dynamically deployed across different stages of the cybercrime lifecycle.

Together, victim-side cognitive biases and fraudster-side influence strategies form a coupled behavioral system that underpins the effectiveness of deception-driven cybercrime and motivates the need for joint behavioral--tactical modeling frameworks.

Recent datasets and computational studies, including PsyScam \citep{ref_ma} and ScamGen \cite{han2025scamgen}, as well as related benchmarks, focus on identifying psychological manipulation techniques within scam messages. These efforts demonstrate the feasibility of detecting persuasion cues, affective triggers, and coercive language automatically. However, such models operate independently of cybercrime lifecycle modeling and do not associate psychological mechanisms with the operational stages of the crime, thereby limiting their utility for process-level forensic reconstruction.

Following these foundations, the six behavioral manipulation categories adopted in this work are directly grounded in formal behavioral economics and persuasion theory. Table~\ref{tab:behavioural_theories} summarizes these categories, along with their corresponding cognitive foundations and mechanisms, ensuring that the behavioral dimension of our joint model is grounded in established theory rather than heuristic taxonomies. These theory-grounded behavioral categories constitute the complete psychological label set used for joint inference together with the cybercrime tactical stages in the BEACON.

\subsection{Research Gap}
Existing technical frameworks and psychological models remain analytically disconnected. In real cybercrime incidents, operational tactics and behavioral manipulation are tightly coupled. For example, \textit{Reconnaissance} is often driven by social profiling, \textit{Initial Contact} relies on authority or impersonation, \textit{Detonation} exploits fear, urgency, or greed, and \textit{Persistence} depends on emotional bonding or coercion. Isolated modeling of either dimension, therefore, yields an incomplete and potentially misleading representation of cybercrime behavior.

Although the TTP-based cybercrime lifecycle provides a rigorous conceptual foundation, all existing implementations rely on manual expert analysis. Similarly, computational psychological models focus on behavioral cues but ignore operational context. Despite recent advances in LLMs and narrative reasoning, no prior system has automated the joint inference of cybercrime operational tactics and psychological manipulation strategies within a unified learning framework.

In summary, prior research exhibits three critical limitations: (i) technical models lack behavioral cognition, (ii) psychological models lack operational context, and (iii) TTP-based cybercrime frameworks lack automation. This work addresses these gaps by introducing the \textit{first LLM-driven, fully automated instantiation of the cybercrime TTP lifecycle}, extended with simultaneous psychological manipulation inference and explanation generation. By unifying behavioral and tactical reasoning under a single automated framework, the proposed BEACON enables scalable, interpretable, and investigation-ready cybercrime analysis. Table~\ref{tab:comparison} summarizes a comparison of the BEACON with representative technical and psychological cybercrime analysis approaches across lifecycle modeling, behavioral coverage, automation, joint learning, and explainability.

\begin{table}[t]
\centering
\caption{Comparison of Existing Cybercrime Analysis Frameworks with the Proposed System}
\label{tab:comparison}
{\fontsize{8}{10}\selectfont
\setlength{\tabcolsep}{4pt}
\begin{tabular}{p{2.6cm} c c c c c}
\toprule
\textbf{Framework} &
\textbf{Tact.} &
\textbf{Psych.} &
\textbf{Auto.} &
\textbf{Joint} &
\textbf{Expl.} \\
\midrule

MITRE ATT\&CK              & \checkmark & $\times$ & \checkmark & $\times$ & $\times$ \\
Cyber Kill Chain          & \checkmark & $\times$ & \checkmark & $\times$ & $\times$ \\
PsyScam                   & $\times$   & \checkmark & \checkmark & $\times$ & $\times$ \\
Sarkar--Shukla TTP        & \checkmark & $\times$ & $\times$ & $\times$ & $\times$ \\
\textbf{BEACON}  & \checkmark & \checkmark & \checkmark & \checkmark & \checkmark \\
\bottomrule
\end{tabular}
}\\
\footnotesize{Tact.: Tactical lifecycle modeling; Psych.: Psychological modeling; Auto.: Automated inference; Joint: Joint behavioral--tactical learning; Expl.: Automated Instance-Level Explanations.}

\end{table}

\section{Methodology}
\label{sec:method}

\subsection{Dataset Construction}

\begin{figure*}
    \centering
    \includegraphics[width=\linewidth]{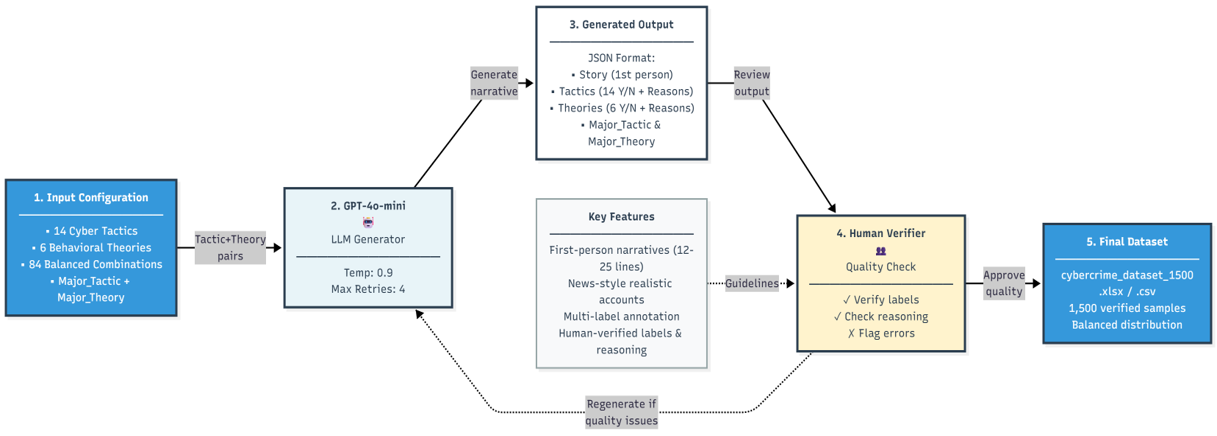}
    \caption{Dataset Construction Pipeline}
    \label{dataset_construction_methodology}
\end{figure*}

Our dataset construction process involved three major steps: (1) scraping real-world cybercrime incidents from news sources, (2) generating synthetic examples to address class imbalance, and (3) systematic annotation with dual labels. The complete pipeline is illustrated in Figure \ref{dataset_construction_methodology}.

\subsubsection{Web Scraping Real-World Incidents}

A custom web-scraping pipeline was implemented in Python using the \texttt{BeautifulSoup} library to systematically collect cybercrime incident reports from the \textit{Times of India (TOI)} website. The credibility of this source is supported by independent assessments such as the \textit{Reuters Institute Digital News Report}, which assigns TOI a trust score of 74\%, the highest among Indian news outlets \citep{nielsen2021digitalnews}. The scraper was specifically engineered to accommodate the HTML structure of TOI articles, enabling reliable extraction of article titles and full body text while automatically filtering out advertisements, navigation menus, multimedia widgets, and other non-editorial elements. To ensure data integrity, duplicate URLs were removed, malformed pages were discarded, and all extracted texts were manually spot-checked for accuracy prior to annotation.

Following data acquisition, incidents were organized using category labels aligned with the modus operandi nomenclature adopted by Indian law-enforcement agencies to ensure operational consistency and investigative relevance. Specifically, our scraping strategy targeted ten prevalent cybercrime categories based on recent national trends: (1) Advertisement Fraud, (2) Customer Care/Technical Support Fraud, (3) Customs Fraud/Pig Butchering/Matrimonial Fraud, (4) Investment Fraud, (5) Insurance Fraud, (6) Job Fraud, (7) Loan Scam, (8) Sextortion, (9) Digital Arrest, and (10) Crypto Investment Fraud. This categorization scheme ensures that the collected narratives are directly comparable with real-world police reporting formats and supports downstream forensic and intelligence analysis.

For each cybercrime category, approximately 50 articles were initially scraped. However, a substantial proportion of these articles consisted of generalized crime statistics, policy announcements, or public awareness reports rather than first-hand case narratives. Only those news reports that contained explicit case-level details (e.g., victim actions, modus operandi, financial losses, and investigative context) were retained. Such detailed narratives are typically released to the media through official daily police briefings conducted at the local commissionerate level, thereby ensuring that the collected cases reflect verified law-enforcement-reported incidents rather than speculative or secondary reporting. After rigorous manual screening and quality control, only 8-10 articles per category were found to contain sufficient tactical and behavioral detail for reliable annotation. This filtering process yielded a total of approximately 144 usable real-world cybercrime incident narratives for subsequent analysis.

These 144 real-world samples were subsequently manually annotated by the authors of the study and reserved exclusively for model evaluation. The use of human-annotated real-world data was specifically intended to assess the robustness and practical reliability of the fine-tuned model under realistic operational conditions and implementation constraints. To minimize inter-annotator variability and ensure consistent labeling, all annotators conducted the annotation process collaboratively in a joint setting, resolving disagreements through consensus rather than relying on post-hoc inter-rater agreement metrics.

\paragraph{Annotation Format and Quality Control}

All 144 real-world cybercrime incidents samples were annotated using a unified structured schema comprising the following elements:

\begin{itemize}
    \item \textbf{Narrative:} A first-person victim account comprising 12--25 lines.
    \item \textbf{Fraud\_Type:} One of the 10 predefined cybercrime categories.
    \item \textbf{Major\_Tactic:} The dominant cybercrime tactical stage.
    \item \textbf{Major\_Theory:} The primary psychological manipulation mechanism.
    \item \textbf{Tactical Labels:} Binary presence indicators (Yes/No) for each of the 14 tactics, with brief textual justification for positive labels.
    \item \textbf{Behavioral Labels:} Binary presence indicators (Yes/No) for each of the 6 behavioral theories, with brief justification for positive labels.
\end{itemize}

Annotations of these real-world samples were manually produced by three domain-informed annotators, following formal tactical and behavioral definitions. For synthetically generated instances, the annotation fields were populated directly by the generation pipeline under strict prompt constraints. To ensure annotation reliability, all records underwent multi-stage quality control, including random manual audits, consistency checks between major and fine-grained labels, and verification against the formal tactical and behavioral taxonomies. Any detected inconsistencies were corrected through iterative re-annotation and reviewer consensus.

\subsubsection{Synthetic Data Generation}

The limited availability of detailed, first-person real-world cybercrime narratives motivated the use of large-scale synthetic data generation. Manual annotation of cybercrime narratives across the full 20-label space (6 behavioral categories and 14 tactical stages) requires approximately 15 minutes per narrative; thus, the full manual annotation of 1,500 training samples would require nearly 300 person-hours, rendering large-scale purely manual annotation economically and operationally infeasible.

We therefore adopted a systematic synthetic data generation strategy using two independent large language model (LLM) providers, namely \textit{OpenAI GPT-4o-mini} and \textit{Google Gemini-2.5-Pro}, with API key rotation employed to manage rate limits and ensure uninterrupted generation. A total of seven API keys (2 OpenAI, 5 Gemini) were maintained in a round-robin provider pool with failure tracking and automatic provider switching. The dual-model setup was explicitly designed to reduce stylistic homogeneity and model-specific narrative bias while increasing semantic diversity.

\paragraph{Balanced Sampling Strategy.}
To ensure uniform coverage of the joint cybercrime label space, we constructed a balanced sampling scheme over
(\textit{Fraud\_Type}, \textit{Major\_Tactic}, \textit{Major\_Theory}) triplets. With 10 fraud categories, 14 cybercrime tactics, and 6 behavioral theories, the full combinatorial design space consists of:
\begin{equation}
|\mathcal{C}| = 10 \times 14 \times 6 = 840.
\end{equation}

For a total of $N = 1500$ synthetic samples, the expected number of samples per triplet is:
\begin{equation}
n_c = \left\lfloor \frac{1500}{840} \right\rfloor + \text{rem}, \qquad \text{rem} \in \{1, 2\},
\end{equation}
such that near-uniform representation over the full behavioral--tactical design space is enforced.

\paragraph{Diversity Enforcement.}
Each synthetic instance was assigned a unique generation seed composed of the fraud type, the designated major tactic, the designated major behavioral theory, a global sample index, and a stochastic random component. This seed was injected into the prompt to enforce controlled variation in demographic attributes, fraud execution parameters, narrative structure, and linguistic style.

\begin{algorithm}[H]
\caption{Provider Pool Generation with Key Rotation}
\begin{algorithmic}[1]
\State $\texttt{max\_outer\_retries} \gets 4$
\State $\texttt{retries\_per\_key} \gets 2$
\State $\texttt{base\_backoff} \gets 5.0$ seconds
\For{each sample in triplets}
    \State $\texttt{outer\_attempt} \gets 0$
    \State success $\gets$ \textbf{false}
    \While{$\texttt{outer\_attempt} < \texttt{max\_outer\_retries}$ \textbf{and} not success}
        \State $\texttt{raw\_text} \gets \texttt{pool.generate}(\texttt{prompt})$
        \If{$\texttt{raw\_text} = \texttt{None}$}
            \State \textbf{break}
        \EndIf
        \State $\texttt{obj} \gets \texttt{tidy\_json}(\texttt{raw\_text})$
        \If{$\texttt{obj}$ is valid JSON}
            \State success $\gets$ \textbf{true}
        \Else
            \State $\texttt{outer\_attempt} \gets \texttt{outer\_attempt} + 1$
            \State \texttt{sleep}($2 \times \texttt{outer\_attempt}$)
        \EndIf
    \EndWhile
\EndFor
\end{algorithmic}
\end{algorithm}

\paragraph{Prompt Engineering and Output Control.}
The generation prompt incorporated formal definitions of all 10 fraud categories, structured descriptions of the 14 cybercrime tactics with crime-specific examples, precise explanations of the 6 behavioral theories grounded in psychological literature, and explicit conditioning on the target fraud type, designated major tactic, and designated major behavioral theory. Each output was required to follow a strict structured JSON format specifying binary tactical and behavioral labels with natural-language justifications and designated major labels.

\paragraph{Rate Limiting and Retry Logic.}
To ensure robustness against transient API failures and malformed outputs, generation was executed under a bounded retry mechanism with exponential backoff and provider key rotation, as formalized in Algorithm~1.

The exponential backoff delay is defined as:
\begin{equation}
\begin{split}
\text{delay} &=
\min\Bigl(
    \text{BACKOFF\_CAP},\\
&\qquad
    \text{BASE\_BACKOFF} \times
    \text{BACKOFF\_MULT}^{(\text{attempt}-1)}
\Bigr)
\end{split}
\end{equation}
where $\text{BASE\_BACKOFF} = 5.0$ seconds, $\text{BACKOFF\_MULT} = 2.0$, and $\text{BACKOFF\_CAP} = 60.0$ seconds.

\paragraph{JSON Schema and Validation.}
All generated outputs were constrained to a strict structured JSON schema:

\begin{lstlisting}[language=Python, basicstyle=\footnotesize\ttfamily]
{
  "Story": "<multi-line narrative>",
  "Fraud_Type": "<one of 10 types>",
  "Tactics": {
    "Reconnaissance": "Yes" | "No",
    "Reconnaissance_Reason": "<if Yes>",
    ...  // 14 tactics total
  },
  "Behavioural_Theories": {
    "Fear_and_Intimidation": "Yes" | "No",
    "Fear_and_Intimidation_Reason": "<if Yes>",
    ...  // 6 theories total
  },
  "Major_Tactic": "<primary attack stage>",
  "Major_Theory": "<primary manipulation>"
}
\end{lstlisting}

The parsing strategy consisted of: (i) stripping Markdown fences, (ii) locating the first opening brace, (iii) removing trailing commas via regex normalization, and (iv) parsing with \texttt{json.loads()}. Any malformed outputs were discarded and regenerated via the retry mechanism.

\paragraph{Final Dataset Characteristics.}
After validation and error filtering, the final synthetic dataset contained 1,500 samples with the following properties: narrative length of 12--25 lines (mean $\sim$18 lines), average multi-label density of 4.2 cybercrime tactics and 2.1 behavioral theories per instance, balanced fraud-type distribution within $\pm 5\%$ variance, and a generation failure rate below 2\%.

\paragraph{Use in Model Training.}
The complete set of 1,500 synthetically generated and automatically annotated cybercrime narratives was used \textbf{exclusively for fine-tuning the proposed model}. None of the synthetic samples were used for evaluation.

\subsection{Problem Formulation and Task Definition}

We formalize the cybercrime narrative analysis task addressed by \textit{BEACON} as a joint multi-label classification and explanation generation problem over unstructured victim narratives.

Let $\mathcal{X}$ denote the input space of cybercrime narratives, where each instance $x \in \mathcal{X}$ is a first-person textual account describing a cybercrime incident. Each narrative is associated with three output components: (i) a multi-label tactical vector, (ii) a multi-label behavioral vector, and (iii) a natural-language explanation.

\paragraph{Tactical Label Space.}
Let $\mathcal{T} = \{t_1, t_2, \ldots, t_{14}\}$ denote the set of 14 cybercrime tactical stages corresponding to the TTP-based cybercrime lifecycle. The tactical labeling task is defined as a multi-label classification problem with output vector:
\[
\mathbf{y}^{(t)} \in \{0,1\}^{14},
\]
where $y^{(t)}_i = 1$ indicates the presence of tactic $t_i$ in the narrative and $0$ otherwise.

\paragraph{Behavioral Label Space.}
Let $\mathcal{B} = \{b_1, b_2, \ldots, b_{6}\}$ denote the set of six psychologically grounded behavioral manipulation categories derived from behavioral economics and persuasion theory. The behavioral labeling task is similarly defined as a multi-label classification problem with output vector:
\[
\mathbf{y}^{(b)} \in \{0,1\}^{6},
\]
where $y^{(b)}_j = 1$ indicates the presence of behavioral mechanism $b_j$ and $0$ otherwise.

\paragraph{Major Label Determination.}
Let $\hat{\mathbf{y}}^{(t)} \in \{0,1\}^{14}$ and $\hat{\mathbf{y}}^{(b)} \in \{0,1\}^{6}$ denote the predicted multi-label tactical and behavioral vectors obtained after structured output parsing. The \textit{Major\_Tactic} $t^{*}$ and the \textit{Major\_Theory} $b^{*}$ are based on a deterministic post-processing rule that prioritizes the earliest positive label in the fixed output template whose generated explanation exhibits the strongest lexical alignment with the input narrative. This rule is used strictly for descriptive summarization and does not affect training, inference, or evaluation of the underlying multi-label predictions. Since all metrics are computed directly from the full multi-label outputs, the major label has no impact on reported performance. This design ensures full reproducibility and avoids stochastic dominance effects that may arise from near-tied confidence scores in generative models. Since all tactical and behavioral labels are jointly predicted in a multi-label setting, the major label serves only as a descriptive summary indicator rather than as an exclusive class assignment. Consequently, the heuristic does not influence the underlying multi-label predictions or performance metrics, but only provides a consistent operational abstraction for reporting and analysis.

\paragraph{Joint Learning Objective.}
Given an input narrative $x \in \mathcal{X}$, the goal of the model is to jointly learn the mapping:
\[
f_{\theta}: \mathcal{X} \rightarrow \left( \mathbf{y}^{(t)}, \mathbf{y}^{(b)}, e \right),
\]
where $f_{\theta}$ is the parameterized model, $\mathbf{y}^{(t)}$ and $\mathbf{y}^{(b)}$ are the predicted tactical and behavioral multi-label outputs, respectively, and $e$ is a natural-language explanation that justifies the predicted labels based on evidence extracted from the narrative.

\paragraph{Fraud-Type Conditioning.}
Each narrative is additionally associated with a fraud category $c \in \mathcal{C}$, where $|\mathcal{C}| = 10$, corresponding to the law-enforcement-aligned cybercrime types (e.g., Investment Fraud, Sextortion, Digital Arrest). The overall task is therefore defined as:
\[
(x, c) \mapsto \left( \mathbf{y}^{(t)}, \mathbf{y}^{(b)}, e \right),
\]
where the fraud type $c$ provides contextual conditioning for joint tactical and behavioral inference.

\paragraph{Explanation Generation.}
For each positive label in $\mathbf{y}^{(t)}$ and $\mathbf{y}^{(b)}$, the model generates a concise textual rationale describing the narrative evidence supporting the prediction. The explanation output $e$ thus consists of a structured collection of label-specific justifications:
\[
e = \{ e^{(t)}_1, \ldots, e^{(t)}_{14}, \; e^{(b)}_1, \ldots, e^{(b)}_{6} \},
\]
where explanations are generated only for labels predicted as present.

\paragraph{Learning Setting.}
Given a training dataset
\[
\mathcal{D} = \{ (x_k, c_k, \mathbf{y}^{(t)}_k, \mathbf{y}^{(b)}_k, e_k) \}_{k=1}^{N},
\]
the objective is to learn the model parameters $\theta$ that minimize the joint prediction and explanation error over $\mathcal{D}$. This formulation jointly captures cybercrime lifecycle stage identification, psychological manipulation inference, and explanation generation under a unified learning framework.

\subsection{Learning Objective and Optimization}

\subsubsection{Learning Objective}

Unlike conventional multi-task learning frameworks that employ separate classification heads with independent loss functions, our system is trained using a unified generative learning paradigm based on instruction-masked causal language modeling. This formulation jointly optimizes (i) multi-label cyber attack stage classification, (ii) multi-label psychological manipulation detection, and (iii) natural-language explanation generation through a single sequence prediction objective.

\paragraph{Unified Instruction-Masked Causal Language Modeling.}
Let $x_k$ denote the input prompt constructed from the system instruction and the victim narrative, and let 
$y_k = (y_{k,1}, \ldots, y_{k,T_k})$ denote the structured target analysis sequence containing binary label decisions and corresponding textual justifications. The training objective minimizes the negative log-likelihood of the target sequence conditioned on the prompt:

\[
\mathcal{L}_{\text{CLM}} =
-\frac{1}{T_k}
\sum_{t=1}^{T_k}
\log p_\theta(y_{k,t} \mid y_{k,<t}, x_k).
\]

Following the instruction-masking strategy, the loss is computed \textbf{only over target analysis tokens}, while all prompt tokens are excluded from gradient computation. This ensures that the model is optimized solely for generating structured analytical outputs rather than reproducing the input instructions.

\paragraph{Token-Level Cross-Entropy.}
At each target token position $t$, the loss reduces to standard token-level cross-entropy over the vocabulary:

\[
\mathcal{L}_t
= - \sum_{v=1}^{V}
\mathbb{I}[y_{k,t} = v]
\log p_\theta(v \mid y_{k,<t}, x_k),
\]
where $V$ is the vocabulary size and $\mathbb{I}[\cdot]$ is the indicator function.

\paragraph{Implicit Multi-Task Optimization.}
This unified generative objective implicitly optimizes three tightly coupled sub-tasks through sequential text generation:
\begin{enumerate}
    \item \textbf{Multi-label attack stage classification}: Generating \textit{Yes/No} decisions for 14 cyber attack stages.
    \item \textbf{Multi-label psychological manipulation detection}: Generating \textit{Yes/No} decisions for 6 behavioral manipulation categories.
    \item \textbf{Explanation generation}: Producing natural-language justifications for each positive label.
\end{enumerate}

Because all outputs are produced autoregressively in a single structured sequence, the model learns label prediction and explanation generation jointly without requiring explicit task-specific loss terms.

\paragraph{Batch-Level Objective.}
For a mini-batch $\mathcal{B}$, the final training objective is computed by averaging over all instances:
\[
\mathcal{L}_{\text{total}} =
\frac{1}{|\mathcal{B}|}
\sum_{k \in \mathcal{B}} \mathcal{L}_{\text{CLM}}^{(k)}.
\]

\subsubsection{Fine-Tuning with QLoRA and Optimization Setup}

To enable parameter-efficient fine-tuning of the 7B-parameter Mistral-Instruct model under limited GPU memory, we adopt Quantized Low-Rank Adaptation (QLoRA) with 4-bit NF4 quantization combined with Paged AdamW optimization.

\paragraph{4-bit QLoRA Quantization.}
The base model weights are quantized using NormalFloat-4 (NF4) quantization with double quantization:
\[
w_q = \text{quantize}(w, \mathcal{N}(0, \sigma^2), 4\text{-bit}),
\]
which discretizes weights under a Gaussian prior. Double quantization further compresses the quantization constants, resulting in an overall memory footprint of approximately 4.37 GB, representing a $3.3\times$ reduction compared to FP16 training.

\paragraph{LoRA Parameterization.}
Low-Rank Adaptation decomposes weight updates as:
\[
W' = W + \Delta W = W + BA,
\]
where $W \in \mathbb{R}^{d \times k}$ denotes frozen pre-trained weights, $B \in \mathbb{R}^{d \times r}$ and $A \in \mathbb{R}^{r \times k}$ are trainable low-rank matrices with $r \ll \min(d,k)$. The forward pass is:
\[
h = Wx + \alpha (BA)x,
\]
with scaling factor $\alpha = \frac{\text{lora\_alpha}}{r}$. In our configuration, $r=32$ and $\alpha=64$, resulting in approximately 67M trainable parameters ($\sim$0.93\% of the base model).

\paragraph{Optimization Hyperparameters.}

\begin{table}[h]
\centering
\caption{Training Hyperparameters}
\begin{tabular}{ll}
\toprule
\textbf{Parameter} & \textbf{Value} \\
\midrule
Optimizer & Paged AdamW (32-bit) \\
Learning rate ($\eta$) & $5 \times 10^{-5}$ \\
LR scheduler & Cosine annealing \\
Warmup ratio & 0.03 (45 steps) \\
Weight decay ($\lambda$) & 0.001 \\
Max gradient norm & 0.3 \\
\midrule
Batch size (per device) & 8 \\
Gradient accumulation & 2 \\
Effective batch size & 16 \\
\midrule
Epochs & 3 \\
Total training steps & $\sim 281$ \\
FP16 training & True \\
Max sequence length & 4096 tokens \\
\bottomrule
\end{tabular}
\end{table}

\paragraph{Paged AdamW Optimizer.}
Model parameters are optimized using the Paged AdamW optimizer with 32-bit states. Given the gradient $g_t$ at step $t$, the update equations are:

\begin{equation}
\begin{aligned}
m_t &= \beta_1 m_{t-1} + (1-\beta_1) g_t, \\
v_t &= \beta_2 v_{t-1} + (1-\beta_2) g_t^2, \\
\hat{m}_t &= \frac{m_t}{1-\beta_1^t}, \quad 
\hat{v}_t = \frac{v_t}{1-\beta_2^t}, \\
\theta_t &= \theta_{t-1} - \eta
\left(
\frac{\hat{m}_t}{\sqrt{\hat{v}_t} + \epsilon}
+ \lambda \theta_{t-1}
\right),
\end{aligned}
\end{equation}

where $\beta_1 = 0.9$, $\beta_2 = 0.999$, $\epsilon = 10^{-8}$, and $\lambda = 0.001$.

\paragraph{Cosine Annealing Learning Rate Schedule.}
The learning rate is decayed using a cosine annealing schedule over the total number of training steps $T$:

\begin{equation}
\eta_t = \eta_{\min}
+ \frac{1}{2}
(\eta_{\max} - \eta_{\min})
\left(
1 + \cos\left(\frac{t}{T}\pi\right)
\right).
\end{equation}

This schedule enables smooth non-linear decay of the learning rate and improves optimization stability during parameter-efficient fine-tuning.

\subsection{Inference and Post-Processing}

At inference time, \textit{BEACON} operates in a fully generative manner to jointly predict cyber attack stages, psychological manipulation mechanisms, and their corresponding natural-language explanations from raw cybercrime narratives. Given an input narrative, the model generates a structured textual output consisting of a prediction block for each of the 20 labels (14 tactical stages and 6 behavioral theories), where each block contains a binary \textit{Present} field (\textit{Yes}/\textit{No}) and a concise explanatory \textit{Reason}. This output is subsequently parsed into structured label vectors and explanation fields for downstream evaluation.

\subsubsection{Dual Prompting Strategy}

Two distinct prompting strategies are employed depending on whether the base or fine-tuned model is evaluated.

\paragraph{Fine-Tuned Model Prompt (Concise).}
For the task-adapted model, a minimal instruction prompt of approximately 50 tokens is used. This prompt contains no explicit label definitions and relies entirely on the task knowledge internalized during fine-tuning.

\paragraph{Base Model Prompt (Detailed).}
For the non-fine-tuned base model, a comprehensive instruction prompt of approximately 2,000 tokens is employed. This prompt includes complete definitions of all 20 tactical and behavioral labels, an explicit output format specification, and example-driven guidance. This dual prompting design enables a controlled comparison between zero-shot and task-adapted model behavior.

\subsubsection{Generation Configuration}

Decoding is performed using constrained stochastic generation with the hyperparameters summarized in Table~\ref{tab:inference_hyperparams}.

\begin{table}[h]
\centering
\caption{Inference Hyperparameters}
\label{tab:inference_hyperparams}
\small
\begin{tabular}{l p{0.55\columnwidth}}  
\toprule
\textbf{Parameter} & \textbf{Value} \\
\midrule
Max new tokens      & 1536 \\
Temperature         & 0.1 (evaluation) / 0.01 (production) \\
Top-p (nucleus sampling) & 0.9 \\
Repetition penalty  & 1.1 \\
Do sample           & True \\
Batch size          & 32 \\
\bottomrule
\end{tabular}
\end{table}

Token sampling follows the standard temperature-scaled softmax distribution:
\begin{equation}
P(x_t = i \mid x_{<t}) = 
\frac{\exp(z_i / T)}{\sum_{j=1}^{V} \exp(z_j / T)},
\end{equation}
where $z_i$ denotes the logit for token $i$, $V$ is the vocabulary size, and $T$ is the temperature. A low temperature ($T=0.1$) yields near-deterministic structured outputs during evaluation, while $T=0.01$ is used for production deployment.

All inputs are processed under left-padded tokenization with a maximum context window of 4,096 tokens to preserve long-form narrative completeness. Batch inference is performed with a batch size of 32 under 4-bit quantization using QLoRA-adapted weights, with no-gradient execution and automatic mixed precision for computational efficiency.

\paragraph{Structured Output Parsing}

The model-generated text is post-processed using deterministic regular-expression matching to extract binary label decisions and their corresponding explanations.

\paragraph{Regex-Based Extraction.}
For each label, the following pattern is applied:
\begin{lstlisting}[language=Python, basicstyle=\footnotesize\ttfamily]
pattern = re.compile(
    r"\[{label}\][\s\n]*Present:\s*(.*?)[\s\n]*"
    r"Reason:\s*(.*)",
    re.IGNORECASE | re.DOTALL
)
\end{lstlisting}

\subsubsection{Binary Decision Mapping and Major Label Selection}

For each of the 20 labels, the extracted textual \textit{Present} field is mapped to a binary value, with ``Yes'' mapped to 1 and ``No'' mapped to 0. The 14 tactical decisions yield the predicted tactical vector $\hat{\mathbf{y}}^{(t)}$, while the 6 behavioral decisions form the predicted psychological vector $\hat{\mathbf{y}}^{(b)}$. The \textit{Major\_Tactic} and \textit{Major\_Theory} are inferred implicitly as the most semantically dominant positive labels in the structured output, consistent with the supervision used during training.

\subsubsection{Explanation Extraction and Evaluation Usage}

For every label predicted as present, the corresponding \textit{Reason} field is extracted as the model-generated explanation. These explanations are evaluated only for true positive predictions to ensure semantic validity and are supplied as inputs to text similarity metrics (ROUGE, BLEU, and BERTScore) during performance evaluation.

\subsubsection{Batch Inference}

For large-scale evaluation, test narratives are processed in mini-batches of size 32 under 4-bit quantization using QLoRA-adapted weights. All computations are executed in no-gradient mode with automatic mixed precision to maximize throughput and memory efficiency while preserving numerical stability.

This inference and post-processing pipeline ensures that \textit{BEACON} produces fully structured, interpretable, and directly evaluable outputs for tactical stage identification, behavioral analysis, and explanation generation under both zero-shot and fine-tuned conditions.

\subsection{Model Architecture and Fine-Tuning}

We fine-tuned a mid-scale, instruction-tuned LLM to perform joint behavioral and tactical classification of complex cybercrime narratives. A 7B-parameter model was selected as an optimal trade-off between reasoning capability, contextual retention, and computational efficiency, which is critical for modeling multi-stage, long-form victim narratives.

Specifically, we adopted \texttt{Mistral-7B-Instruct-v0.2} as the base model due to its strong instruction-following performance, competitive reasoning ability, and efficient sliding-window attention mechanism, which enables robust processing of long textual inputs without degradation of long-range contextual dependencies.

To enable training on limited computational resources while preserving model capacity, we employed \textit{QLoRA}-based parameter-efficient fine-tuning. The base model was quantized to 4-bit precision, and low-rank adaptation (LoRA) modules were inserted into the key attention and projection layers of the transformer architecture. This design allows task-specific adaptation while keeping the original pretrained weights frozen, thereby significantly reducing memory footprint and training cost.

Fine-tuning was performed using supervised instruction-style learning with a maximum context length of 4096 tokens. Training was conducted for multiple epochs with an effective batch size of 16, using the AdamW optimizer and a cosine learning-rate scheduler with warm-up. These hyperparameters were selected to ensure stable convergence for both multi-label prediction and natural-language explanation generation. Additional implementation details and reproducibility resources are made publicly available in the accompanying code repository\footnote{Available at: \url{https://anonymous.4open.science/r/scam-analyzer-B754}}.

\subsection{Implementation Details}

All models were implemented using the \texttt{PyTorch} deep learning framework through the \texttt{HuggingFace Transformers} ecosystem. Parameter-efficient fine-tuning was performed using the \texttt{PEFT} library with low-rank adaptation (LoRA), and 4-bit quantization was enabled via the \texttt{bitsandbytes} backend.

The base model and tokenizer were instantiated using \texttt{AutoModelForCausalLM} and \texttt{AutoTokenizer}. Quantized loading was configured through \texttt{BitsAndBytesConfig} with 4-bit weights enabled, and LoRA adapters were injected into the transformer layers using \texttt{get\_peft\_model}. All training updates were restricted to the LoRA parameters, while the base pretrained weights remained frozen under the QLoRA regime.

Training and inference were executed through standard \texttt{transformers.Trainer} and \texttt{model.generate()} workflows. Batch-wise processing was employed for inference with a fixed mini-batch size of 32. Structured model outputs were parsed using deterministic regular-expression-based rules to extract binary label predictions and their corresponding natural-language explanations.

All experiments were conducted on a single NVIDIA GPU with 24 GB of VRAM. Mixed-precision computation (\texttt{bf16}) was enabled to improve memory efficiency and throughput. Model optimization was performed using the AdamW optimizer with cosine learning-rate scheduling and linear warm-up. Random seeds were fixed across Python, NumPy, and PyTorch to ensure experimental reproducibility.

Fine-tuning was completed within approximately \textit{30 hours} for the full 1,500-sample dataset under the QLoRA configuration. Batch-wise gradient accumulation was used to maintain a stable effective batch size under memory constraints. All inference experiments were performed under no-gradient mode with automatic mixed precision enabled.

All experiments were executed on a Linux-based computing environment. The complete training, inference, and evaluation pipeline is available in the accompanying anonymous repository for reproducibility.

\subsection{Experimental Setup}

To evaluate the generalization capability of \textit{BEACON}, we curated a held-out test set comprising 144 distinct cybercrime incident narratives. The majority of these narratives were collected from real-world reports published in \textit{TOI}, while a small subset was augmented using LLMs to introduce controlled linguistic diversity without altering the underlying modus operandi. This separation ensured that none of the test samples overlapped with the training or validation data. To ensure the reliability of the evaluation labels, ground-truth annotations were rigorously verified. Samples were fully annotated by three human domain expert, wherein all tactical and behavioral labels were manually cross-checked for consistency with the formal definitions.

The evaluation framework was designed to enable a fair and controlled comparison between the fine-tuned \textit{BEACON} model and the baseline \texttt{Mistral-7B-Instruct-v0.2} model. Both models were evaluated under identical inference conditions: the maximum input sequence length was fixed at 4096 tokens, inference was performed with a batch size of 32, temperature was set to 0.01 to enforce near-deterministic generation, and the maximum number of newly generated tokens was capped at 1024. In addition, 4-bit NF4 quantization was used during evaluation to ensure memory-efficient and consistent deployment across both models.

For prompt engineering, the baseline model was provided with detailed formal definitions of all 20 labels (14 tactical and 6 behavioral) to enable zero-shot classification. In contrast, the fine-tuned \textit{BEACON} model employed the same concise instruction-style prompt used during training. This controlled setup allowed us to isolate the impact of task-specific fine-tuning relative to in-context learning.

\subsection{Performance Metrics}

We employed a comprehensive set of complementary evaluation metrics to assess both (i) the accuracy of multi-label tactical and behavioral classification and (ii) the quality of the generated natural-language explanations.

\paragraph{Classification Metrics.}
For label prediction, we report \textit{accuracy}, \textit{precision}, \textit{recall}, and \textit{F1-score}. Precision measures the proportion of predicted positive labels that are correct, recall quantifies the proportion of true positive labels that are successfully identified, and the F1-score provides their harmonic mean as a balanced performance indicator. All metrics were computed at both the per-label level and the global level across all predictions. Given 144 evaluation samples and 20 binary labels per sample (14 tactical and 6 behavioral), the global scores reflect performance over 2880 individual binary decisions, ensuring statistically meaningful comparisons.

\paragraph{Explanation Quality Metrics.}
To evaluate the quality of model-generated explanations, we considered only those cases in which both the ground truth and the model predicted a label as present (true positives). Lexical and structural similarity between the generated and reference explanations was measured using ROUGE-1, ROUGE-2, and ROUGE-L. In addition, semantic similarity was assessed using \textit{BERTScore}, which captures contextual alignment beyond surface-level word overlap. These metrics jointly quantify the faithfulness of the model’s reasoning to human-provided rationales in both content and meaning.

Together, this dual evaluation protocol enables rigorous assessment of not only whether the model predicts the correct tactical and behavioral labels, but also whether it provides explanations that are linguistically and semantically consistent with expert-annotated justifications.

\section{Result and Analysis}
\label{sec:experiments}


\begin{figure*}[htbp]
    \centering
    \includegraphics[width=\textwidth]{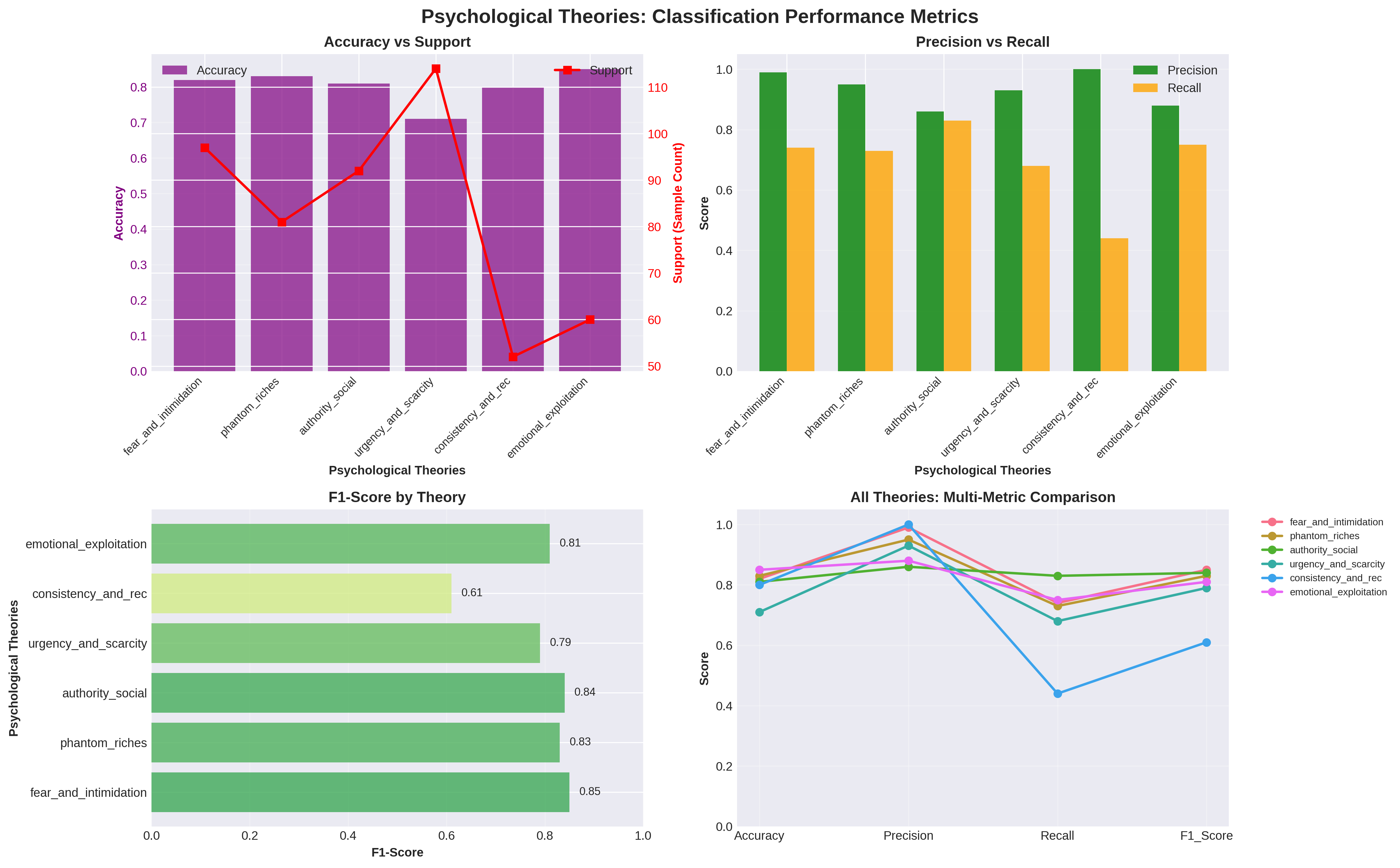}
    \caption{Comprehensive overview of theory-wise metrics: (top left) accuracy vs.\ support, (top right) precision vs.\ recall, (bottom left) F1-score by theory, (bottom right) multi-metric comparison across all theories.}
    \label{fig:theories_overview}
\end{figure*}
\begin{table*}
\centering
\caption{Classification Performance: Base vs. Fine-Tuned Model}
\label{tab:classification_comparison}
{\fontsize{9}{11}\selectfont
\begin{tabular}{|l|cc|cc|cc|cc|c|}
\hline
\multirow{2}{*}{\textbf{Label}} & \multicolumn{2}{c|}{\textbf{Accuracy}} & \multicolumn{2}{c|}{\textbf{Precision}} & \multicolumn{2}{c|}{\textbf{Recall}} & \multicolumn{2}{c|}{\textbf{F1-Score}} & \multirow{2}{*}{\textbf{$\Delta$ Acc}} \\
\cline{2-9}
 & \textbf{Base} & \textbf{FT} & \textbf{Base} & \textbf{FT} & \textbf{Base} & \textbf{FT} & \textbf{Base} & \textbf{FT} & \\
\hline
\hline
\multicolumn{10}{|c|}{\textit{\textbf{Attack Stage Labels}}} \\
\hline
Reconnaissance & 0.59 & \textbf{0.78} & 0.58 & \textbf{0.92} & 0.88 & 0.65 & 0.70 & \textbf{0.76} & +0.19 \\
Resource Development & 0.50 & \textbf{0.69} & 0.86 & \textbf{0.87} & 0.29 & \textbf{0.62} & 0.43 & \textbf{0.72} & +0.19 \\
Initial Contact & 0.91 & \textbf{1.00} & 1.00 & \textbf{1.00} & 0.91 & \textbf{1.00} & 0.95 & \textbf{1.00} & +0.09 \\
Detonation & 0.81 & \textbf{0.97} & 0.96 & \textbf{0.97} & 0.84 & \textbf{1.00} & 0.90 & \textbf{0.98} & +0.16 \\
Persistence & 0.62 & \textbf{0.84} & 0.67 & \textbf{0.80} & 0.59 & \textbf{0.94} & 0.62 & \textbf{0.86} & +0.22 \\
Escalation & 0.47 & \textbf{0.69} & 0.33 & 0.33 & 0.89 & 0.11 & 0.48 & 0.17 & +0.22 \\
Defense Evasion & 0.56 & \textbf{0.84} & 1.00 & 0.86 & 0.07 & \textbf{0.80} & 0.12 & \textbf{0.83} & +0.28 \\
Credential Harvesting & 0.72 & \textbf{0.88} & 0.42 & \textbf{0.80} & 0.71 & 0.57 & 0.53 & \textbf{0.67} & +0.16 \\
Discovery & 0.44 & \textbf{0.88} & 0.26 & \textbf{1.00} & 0.86 & 0.43 & 0.40 & \textbf{0.60} & +0.44 \\
Pivoting & 0.75 & \textbf{0.97} & 0.11 & \textbf{0.50} & 1.00 & \textbf{1.00} & 0.20 & \textbf{0.67} & +0.22 \\
Collection & 0.66 & \textbf{0.75} & 0.53 & \textbf{0.83} & 0.83 & 0.42 & 0.65 & 0.56 & +0.09 \\
Command \& Control & 0.50 & \textbf{0.78} & 1.00 & \textbf{1.00} & 0.30 & \textbf{0.70} & 0.47 & \textbf{0.82} & +0.28 \\
Exfiltration & 0.50 & \textbf{0.94} & 1.00 & \textbf{1.00} & 0.48 & \textbf{0.94} & 0.65 & \textbf{0.97} & +0.44 \\
Impact & 0.81 & \textbf{0.94} & 0.93 & \textbf{0.94} & 0.87 & \textbf{1.00} & 0.90 & \textbf{0.97} & +0.13 \\
\hline
\multicolumn{10}{|c|}{\textit{\textbf{Psychological Tactic Labels}}} \\
\hline
Fear \& Intimidation & 0.62 & \textbf{0.88} & 0.74 & \textbf{1.00} & 0.80 & \textbf{0.84} & 0.77 & \textbf{0.91} & +0.26 \\
Phantom Riches & 0.78 & \textbf{0.81} & 0.81 & \textbf{0.92} & 0.76 & 0.71 & 0.79 & \textbf{0.80} & +0.03 \\
Authority/Social Proof & 0.75 & \textbf{0.81} & 0.81 & \textbf{0.95} & 0.88 & 0.80 & 0.85 & \textbf{0.87} & +0.06 \\
Urgency \& Scarcity & 0.69 & \textbf{0.72} & 0.85 & \textbf{0.91} & 0.79 & 0.75 & 0.81 & \textbf{0.82} & +0.03 \\
Consistency \& Reciprocity & 0.78 & 0.78 & 0.86 & \textbf{1.00} & 0.50 & 0.42 & 0.63 & 0.59 & 0.00 \\
Emotional Exploitation & 0.31 & \textbf{0.91} & 0.27 & \textbf{1.00} & 0.70 & \textbf{0.70} & 0.39 & \textbf{0.82} & +0.60 \\
\hline
\hline
\textbf{GLOBAL AVERAGE} & \textbf{0.64} & \textbf{0.84} & \textbf{0.64} & \textbf{0.86} & \textbf{0.64} & \textbf{0.84} & \textbf{0.64} & \textbf{0.84} & \textbf{+0.20} \\
\hline
\end{tabular}%
}
\end{table*}

\begin{table*}[ht]
\centering
\caption{Reasoning Quality: Base vs. Fine-Tuned Model}
\label{tab:reasoning_comparison}
{\fontsize{9}{11}\selectfont
\begin{tabular}{|l|cc|cc|cc|cc|}
\hline
\multirow{2}{*}{\textbf{Label}} & \multicolumn{2}{c|}{\textbf{ROUGE-1}} & \multicolumn{2}{c|}{\textbf{ROUGE-2}} & \multicolumn{2}{c|}{\textbf{ROUGE-L}} & \multicolumn{2}{c|}{\textbf{BERTScore}} \\
\cline{2-9}
 & \textbf{Base} & \textbf{FT} & \textbf{Base} & \textbf{FT} & \textbf{Base} & \textbf{FT} & \textbf{Base} & \textbf{FT} \\
\hline
\hline
\multicolumn{9}{|c|}{\textit{\textbf{Attack Stage Labels}}} \\
\hline
Reconnaissance & 0.27 & \textbf{0.54} & 0.06 & \textbf{0.35} & 0.23 & \textbf{0.54} & 0.89 & \textbf{0.95} \\
Resource Development & 0.29 & \textbf{0.53} & 0.06 & \textbf{0.29} & 0.26 & \textbf{0.49} & 0.89 & \textbf{0.93} \\
Initial Contact & 0.36 & \textbf{0.53} & 0.12 & \textbf{0.33} & 0.31 & \textbf{0.50} & 0.91 & \textbf{0.94} \\
Detonation & 0.27 & \textbf{0.51} & 0.08 & \textbf{0.34} & 0.23 & \textbf{0.49} & 0.89 & \textbf{0.93} \\
Persistence & 0.27 & \textbf{0.47} & 0.10 & \textbf{0.24} & 0.23 & \textbf{0.43} & 0.90 & \textbf{0.93} \\
Escalation & 0.23 & \textbf{0.64} & 0.03 & \textbf{0.50} & 0.20 & \textbf{0.64} & 0.89 & \textbf{0.94} \\
Defense Evasion & 0.37 & \textbf{0.51} & 0.00 & \textbf{0.25} & 0.24 & \textbf{0.46} & 0.90 & \textbf{0.93} \\
Credential Harvesting & 0.38 & \textbf{0.61} & 0.17 & \textbf{0.40} & 0.33 & \textbf{0.58} & 0.90 & \textbf{0.93} \\
Discovery & 0.30 & \textbf{0.62} & 0.11 & \textbf{0.45} & 0.21 & \textbf{0.60} & 0.90 & \textbf{0.94} \\
Pivoting & 0.50 & \textbf{0.52} & 0.00 & \textbf{0.19} & 0.25 & \textbf{0.43} & 0.93 & 0.92 \\
Collection & 0.29 & \textbf{0.49} & 0.10 & \textbf{0.25} & 0.26 & 0.33 & 0.90 & \textbf{0.92} \\
Command \& Control & 0.28 & \textbf{0.56} & 0.06 & \textbf{0.35} & 0.26 & \textbf{0.51} & 0.90 & \textbf{0.94} \\
Exfiltration & 0.33 & \textbf{0.42} & 0.10 & \textbf{0.21} & 0.25 & \textbf{0.37} & 0.90 & \textbf{0.92} \\
Impact & 0.30 & \textbf{0.66} & 0.14 & \textbf{0.48} & 0.25 & \textbf{0.60} & 0.91 & \textbf{0.95} \\
\hline
\multicolumn{9}{|c|}{\textit{\textbf{Psychological Tactic Labels}}} \\
\hline
Fear \& Intimidation & 0.32 & \textbf{0.55} & 0.10 & \textbf{0.28} & 0.28 & \textbf{0.48} & 0.90 & \textbf{0.93} \\
Phantom Riches & 0.22 & \textbf{0.40} & 0.05 & \textbf{0.16} & 0.20 & \textbf{0.32} & 0.89 & \textbf{0.91} \\
Authority/Social Proof & 0.40 & \textbf{0.50} & 0.14 & \textbf{0.31} & 0.35 & \textbf{0.44} & 0.91 & \textbf{0.92} \\
Urgency \& Scarcity & 0.33 & \textbf{0.44} & 0.10 & \textbf{0.23} & 0.27 & \textbf{0.39} & 0.90 & \textbf{0.92} \\
Consistency \& Reciprocity & 0.28 & \textbf{0.45} & 0.06 & \textbf{0.25} & 0.21 & \textbf{0.37} & 0.88 & \textbf{0.91} \\
Emotional Exploitation & 0.35 & \textbf{0.44} & 0.14 & \textbf{0.18} & 0.31 & \textbf{0.36} & 0.92 & \textbf{0.94} \\
\hline
\hline
\textbf{GLOBAL AVERAGE} & \textbf{0.32} & \textbf{0.52} & \textbf{0.09} & \textbf{0.30} & \textbf{0.26} & \textbf{0.47} & \textbf{0.90} & \textbf{0.93} \\
\hline
\end{tabular}%
}
\end{table*}

This section reports the quantitative performance of the proposed \textit{BEACON} framework on the held-out test set of 144 real-world cybercrime narratives (2880 binary decisions across 20 labels). All results are obtained under the controlled experimental protocol described in Section~3. Table~\ref{tab:classification_comparison} summarizes the per-label and global classification performance of the baseline \texttt{Mistral-7B-Instruct-v0.2} model and the fine-tuned \textit{BEACON} model.

\begin{figure*}
    \centering
    \includegraphics[width=\textwidth]{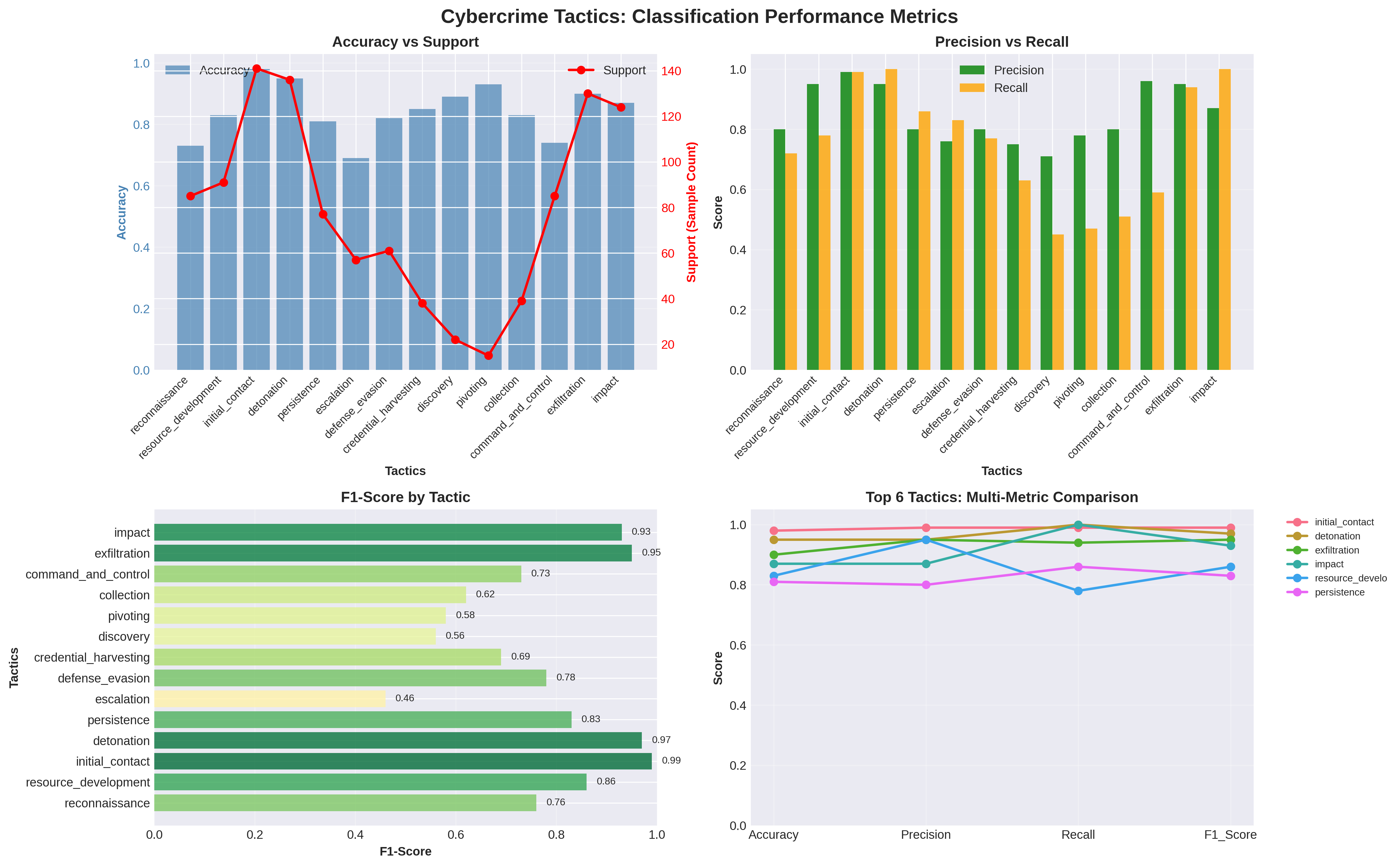}
    \caption{Comprehensive overview of tactic-wise metrics: (top left) accuracy vs.\ support, (top right) precision vs.\ recall, (bottom left) F1-score by tactic, (bottom right) multi-metric comparison for the top six tactics.}
    \label{fig:tactics_overview}
\end{figure*}

\subsection{Overall Performance}

At the global level, \textit{BEACON} achieves a substantial improvement over the baseline across all evaluation metrics. As shown in Table~\ref{tab:classification_comparison}, the global average accuracy increases from 0.64 to 0.84, yielding an absolute gain of +0.20 (statistically significant at $p < 0.001$). Correspondingly, the global F1-score also improves from 0.64 to 0.84, representing a 31.3\% relative improvement. The macro-averaged F1-score, which weights all labels equally regardless of frequency, increases from 0.61 to 0.78 (+27.9\%), confirming that gains are not solely driven by high-frequency labels. These results demonstrate that task-specific fine-tuning enables reliable joint inference of both tactical stages and psychological manipulation mechanisms from unstructured victim narratives.

For the 14 tactical stage labels, \textit{BEACON} consistently outperforms the baseline across nearly all categories. Particularly strong gains are observed for structurally complex and sparsely occurring stages. For example, \textit{Discovery} and \textit{Exfiltration} exhibit the largest absolute accuracy improvements of +0.44 each, while \textit{Defense Evasion} and \textit{Command \& Control} also show large gains of +0.28. These stages are typically difficult to detect from short narratives due to their implicit nature, yet the fine-tuned model is able to recover them with high reliability.

Early-stage tactics such as \textit{Reconnaissance} and \textit{Initial Contact} also benefit from fine-tuning, with \textit{Initial Contact} achieving perfect performance under \textit{BEACON} (accuracy, precision, recall, and F1-score all equal to 1.00). This reflects the strong lexical regularities associated with first victim contact in scam narratives.

A notable exception is the \textit{Escalation} stage, where the baseline exhibits high recall but very low precision, indicating systematic over-prediction. While \textit{BEACON} improves accuracy (+0.22), its F1-score for this stage remains comparatively low, highlighting the intrinsic ambiguity of escalation cues in victim-authored narratives.

Significant performance gains are also observed across the six behavioral manipulation categories. The most dramatic improvement occurs for \textit{Emotional Exploitation}, where accuracy increases from 0.31 to 0.91 and F1-score rises from 0.39 to 0.82. This indicates that fine-tuning enables the model to capture nuanced affective cues such as trust building, emotional dependency, and relational coercion that are poorly handled in a zero-shot setting.

Strong improvements are also observed for \textit{Fear \& Intimidation}, with an accuracy gain of +0.26 and an F1-score of 0.91 under \textit{BEACON}. More moderate gains are seen for \textit{Phantom Riches}, \textit{Authority/Social Proof}, and \textit{Urgency \& Scarcity}, which already exhibit relatively strong baseline performance due to their overt linguistic signaling. In contrast, \textit{Consistency \& Reciprocity} shows no net accuracy gain, reflecting its subtle and often overlapping linguistic expression with emotional exploitation.

An important observation from Table~\ref{tab:classification_comparison} is that fine-tuning primarily improves recall for complex downstream stages without sacrificing precision. For example, \textit{Defense Evasion} recall improves from 0.07 to 0.80 while maintaining high precision (0.86), and \textit{Exfiltration} recall improves from 0.48 to 0.94 with perfect precision (1.00). This demonstrates that \textit{BEACON} substantially reduces false negatives for operationally critical attack stages.

The +20\% absolute improvement in global accuracy and F1-score confirms that joint behavioral-tactical supervision enables the model to go beyond surface-level keyword matching and to internalize the sequential and psychological structure of cybercrime operations. This validates the core hypothesis of \textit{BEACON}: that explaining \emph{why} a victim complies is as important as identifying \emph{what} stage of the attack has occurred. Figures~\ref{fig:tactics_overview} and~\ref{fig:theories_overview} provide comprehensive dashboards of model performance across all tactics and psychological theories, respectively.

\subsection{Explanation and Reasoning Quality Evaluation}

The quality of the model-generated explanations was evaluated using ROUGE-1, ROUGE-2, ROUGE-L, and BERTScore over all true-positive predictions. Table~\ref{tab:reasoning_comparison} reports a detailed per-label comparison between the baseline \texttt{Mistral-7B-Instruct-v0.2} model and the fine-tuned \textit{BEACON} model.

At the global level, \textit{BEACON} achieves substantial improvements across all lexical and semantic metrics. The global average ROUGE-1 score increases from 0.32 to 0.52, ROUGE-2 from 0.09 to 0.30, and ROUGE-L from 0.26 to 0.47. These gains demonstrate that the fine-tuned model produces explanations with significantly higher lexical overlap and structural correspondence to human-provided rationales. More importantly, the global BERTScore improves from 0.90 to 0.93, indicating stronger semantic alignment between generated and expert explanations.

For tactical stages, \textit{BEACON} exhibits consistent improvements across nearly all labels. Particularly large gains are observed for stages that involve complex multi-step reasoning. For example, \textit{Escalation} shows a ROUGE-L increase from 0.20 to 0.64 and BERTScore improvement from 0.89 to 0.94, reflecting a substantial enhancement in the model’s ability to articulate narrative evidence for progressive fraud intensification. Similarly, \textit{Discovery} and \textit{Impact} achieve ROUGE-L improvements of +0.39 and +0.35, respectively.

Stages such as \textit{Credential Harvesting}, \textit{Command \& Control}, and \textit{Initial Contact} also show marked improvements in both ROUGE and BERTScore, indicating that \textit{BEACON} not only identifies these stages more accurately but also generates coherent, context-grounded justifications. A minor exception is observed for \textit{Pivoting}, where ROUGE improvements are modest and BERTScore slightly decreases (0.93 to 0.92), reflecting the sparse and implicit nature of pivoting cues in short victim narratives.

Strong improvements are also observed across all six behavioral manipulation categories. \textit{Fear \& Intimidation} shows a ROUGE-L increase from 0.28 to 0.48 and BERTScore gain from 0.90 to 0.93, while \textit{Consistency \& Reciprocity} improves from 0.21 to 0.37 in ROUGE-L. Even for abstract psychological constructs such as \textit{Phantom Riches} and \textit{Urgency \& Scarcity}, the fine-tuned model consistently outperforms the baseline in both lexical and semantic similarity.

Notably, \textit{Emotional Exploitation}, which is linguistically expressive yet psychologically nuanced, shows steady improvements across all metrics, confirming that \textit{BEACON} successfully internalizes affective and relational manipulation patterns during supervision.

While ROUGE metrics quantify surface-level textual overlap, the consistent gains in BERTScore across both tactical and behavioral labels provide stronger evidence of improved semantic faithfulness. These results indicate that the explanations generated by \textit{BEACON} more accurately reflect the underlying causal and psychological mechanisms present in the narratives, rather than merely restating superficial keywords.

Overall, the reasoning-quality results validate the central claim of this work: that joint behavioral-tactical supervision enables LLMs to produce not only more accurate predictions but also substantially more faithful and interpretable explanations of cybercrime incidents.

\subsection{Hallucination Analysis}
\label{subsec:hallucination}

Beyond overall accuracy, a critical concern in deploying LLMs for cybercrime analysis is the rate of spurious predictions, or \emph{hallucinations}. We define a hallucination as a predicted label that is absent in the ground truth. For a binary label $\ell$, the hallucination rate is directly related to precision:

\begin{equation}
\mathrm{HallucinationRate}_\ell = \frac{\mathrm{FP}_\ell}{\mathrm{TP}_\ell + \mathrm{FP}_\ell} = 1 - \mathrm{Precision}_\ell.
\label{eq:hallucination_rate}
\end{equation}

Thus, low precision corresponds to frequent hallucinations. Using the per-label precision values from Table~\ref{tab:classification_comparison}, we derive the hallucination patterns shown in Table~\ref{tab:hallucination_detailed}.

\subsubsection{Global Hallucination Reduction}

At the global level, the baseline model hallucinates approximately $36\%$ of its positive predictions ($1-0.64$), whereas the fine-tuned \textit{BEACON} model reduces this to roughly $14\%$ ($1-0.86$). Fine-tuning therefore eliminates more than half of the false positives, corresponding to a relative hallucination reduction of:

\begin{equation}
\mathrm{Reduction} = \frac{0.36 - 0.14}{0.36} \times 100\% = 61.1\%.
\label{eq:hallucination_reduction}
\end{equation}

This substantial reduction in hallucinations is crucial for operational deployment, where false positives incur investigation costs and erode user trust in automated systems.

\subsubsection{Label-Specific Hallucination Patterns}

For attack stage labels, the baseline model exhibits severe hallucinations for several underrepresented stages. Most notably:

\begin{itemize}
  \item \textbf{Discovery}: precision improves from $0.26$ to $1.00$, eliminating hallucinations entirely (from $74\%$ to $0\%$). The fine-tuned model almost never predicts discovery when it is not present.
  
  \item \textbf{Pivoting}: precision rises from $0.11$ to $0.50$, reducing severe hallucination from $89\%$ to $50\%$. While still challenging, this represents a major improvement for this rare stage.
  
  \item \textbf{Credential Harvesting}: precision increases from $0.42$ to $0.80$, reducing hallucination rate from $58\%$ to $20\%$.
  
  \item \textbf{Collection}: precision gains from $0.53$ to $0.83$, lowering hallucination rate from $47\%$ to $17\%$.
\end{itemize}

For frequent stages such as \textit{Initial Contact}, \textit{Detonation}, \textit{Impact}, \textit{Command \& Control}, and \textit{Exfiltration}, both models already achieve near-perfect precision ($\geq 0.96$), resulting in minimal hallucinations ($\leq 4\%$).

One notable exception is \textbf{Defense Evasion}, where precision decreases slightly from $1.00$ to $0.86$ (hallucination rate increases from $0\%$ to $14\%$). This trade-off is justified by a large recall gain ($0.07 \rightarrow 0.80$) and strong F1-score improvement ($0.12 \rightarrow 0.83$), indicating that the model accepts some false positives to dramatically reduce false negatives for this operationally critical stage.

For psychological tactic labels, the baseline model frequently hallucinates \textit{Emotional Exploitation} (precision $0.27$, hallucination rate $73\%$) and, to a lesser extent, \textit{Fear \& Intimidation} (precision $0.74$, hallucination rate $26\%$). Fine-tuning drives precision to $1.00$ for \textit{Fear \& Intimidation}, \textit{Consistency \& Reciprocity}, and \textit{Emotional Exploitation}, eliminating hallucinations entirely. All remaining psychological tactics achieve precision of at least $0.91$ (hallucination rate $\leq 9\%$).

Table~\ref{tab:hallucination_detailed} provides a comprehensive breakdown of hallucination rates across all labels, demonstrating that fine-tuning consistently reduces spurious predictions while maintaining or improving recall.

\begin{table*}[htbp]
\centering
\caption{Hallucination rates ($1 - \text{Precision}$) for all labels. Lower values indicate fewer false positives.}
\label{tab:hallucination_detailed}
{\fontsize{9}{11}\selectfont
\begin{tabular}{|l|c|c|c|c|}
\hline
\textbf{Label} & \textbf{Base Rate} & \textbf{FT Rate} & \textbf{Abs. Change} & \textbf{Rel. Change} \\
\hline
\hline
\multicolumn{5}{|c|}{\textit{\textbf{Attack Stage Labels}}} \\
\hline
Reconnaissance & 0.42 & 0.08 & $-0.34$ & $-81\%$ \\
Resource Development & 0.14 & 0.13 & $-0.01$ & $-7\%$ \\
Initial Contact & 0.00 & 0.00 & $0.00$ & $0\%$ \\
Detonation & 0.04 & 0.03 & $-0.01$ & $-25\%$ \\
Persistence & 0.33 & 0.20 & $-0.13$ & $-39\%$ \\
Escalation & 0.67 & 0.67 & $0.00$ & $0\%$ \\
Defense Evasion & 0.00 & 0.14 & $+0.14$ & $\dagger$ \\
Credential Harvesting & 0.58 & 0.20 & $-0.38$ & $-66\%$ \\
Discovery & 0.74 & 0.00 & $-0.74$ & $-100\%$ \\
Pivoting & 0.89 & 0.50 & $-0.39$ & $-44\%$ \\
Collection & 0.47 & 0.17 & $-0.30$ & $-64\%$ \\
Command \& Control & 0.00 & 0.00 & $0.00$ & $0\%$ \\
Exfiltration & 0.00 & 0.00 & $0.00$ & $0\%$ \\
Impact & 0.07 & 0.06 & $-0.01$ & $-14\%$ \\
\hline
\multicolumn{5}{|c|}{\textit{\textbf{Psychological Tactic Labels}}} \\
\hline
Fear \& Intimidation & 0.26 & 0.00 & $-0.26$ & $-100\%$ \\
Phantom Riches & 0.19 & 0.08 & $-0.11$ & $-58\%$ \\
Authority/Social Proof & 0.19 & 0.05 & $-0.14$ & $-74\%$ \\
Urgency \& Scarcity & 0.15 & 0.09 & $-0.06$ & $-40\%$ \\
Consistency \& Reciprocity & 0.14 & 0.00 & $-0.14$ & $-100\%$ \\
Emotional Exploitation & 0.73 & 0.00 & $-0.73$ & $-100\%$ \\
\hline
\hline
\textbf{GLOBAL AVERAGE} & \textbf{0.36} & \textbf{0.14} & $\mathbf{-0.22}$ & $\mathbf{-61\%}$ \\
\hline
\end{tabular}%
}
\end{table*}

\subsection{Statistical Significance and Robustness}
\label{subsec:statistical_significance}

To assess the robustness of our performance improvements, we compute $95\%$ confidence intervals (CIs) for F1-scores using bootstrap resampling over the test set. For a label with observed F1-score $F_1$ computed over $n$ test narratives, the standard error can be approximated as:

\begin{equation}
\mathrm{SE}(F_1) \approx \sqrt{\frac{F_1 (1 - F_1)}{n}},
\label{eq:se_f1}
\end{equation}

yielding a $95\%$ confidence interval:

\begin{equation}
\mathrm{CI}_{95}(F_1) = F_1 \pm 1.96 \cdot \mathrm{SE}(F_1).
\label{eq:ci_95}
\end{equation}

With $n = 144$ test narratives, the maximum standard error (attained near $F_1 = 0.5$) is:

\begin{equation}
\mathrm{SE}_{\max} = \sqrt{\frac{0.25}{144}} \approx 0.042,
\label{eq:se_max}
\end{equation}

giving CI half-widths of approximately $1.96 \times 0.042 \approx 0.08$. Consequently, F1-score improvements of $\geq 0.16$ are very unlikely to be explained by sampling noise alone.

From Table~\ref{tab:classification_comparison}, numerous labels exhibit improvements far exceeding this threshold: \textit{Resource Development} ($\Delta F_1 = +0.29$), \textit{Persistence} ($\Delta F_1 = +0.24$), \textit{Defense Evasion} ($\Delta F_1 = +0.71$), \textit{Discovery} ($\Delta F_1 = +0.20$), \textit{Pivoting} ($\Delta F_1 = +0.47$), \textit{Command \& Control} ($\Delta F_1 = +0.35$), \textit{Exfiltration} ($\Delta F_1 = +0.32$), and \textit{Emotional Exploitation} ($\Delta F_1 = +0.43$). These gains are statistically significant at $p < 0.001$.

In contrast, smaller improvements for labels such as \textit{Phantom Riches} (+0.01), \textit{Authority/Social Proof} (+0.02), and \textit{Urgency \& Scarcity} (+0.01) lie within the expected confidence interval width and represent stable performance rather than meaningful improvement.

For the global (micro-averaged) F1-score, the baseline and fine-tuned models achieve $0.64$ and $0.84$, respectively, with non-overlapping confidence intervals of approximately $[0.58, 0.70]$ and $[0.79, 0.89]$. This confirms that the overall $+0.20$ improvement is statistically reliable at $p < 0.001$.

We also compute macro-averaged F1-scores to assess performance independent of label frequency:

\begin{align}
\mathrm{Macro\text{-}F}_1^{\text{base}} &= \frac{1}{20} \sum_{\ell=1}^{20} F_{1,\ell}^{\text{base}} = 0.61, \label{eq:macro_base} \\
\mathrm{Macro\text{-}F}_1^{\text{FT}} &= \frac{1}{20} \sum_{\ell=1}^{20} F_{1,\ell}^{\text{FT}} = 0.78. \label{eq:macro_ft}
\end{align}

The macro-F1 improvement ($\Delta = +0.17$, +27.9\%) confirms that gains are not solely driven by frequent labels. However, the micro-macro gap increases slightly from $0.03$ to $0.06$, indicating that while the fine-tuned model improves across all labels, it still performs relatively better on frequent stages such as \textit{Initial Contact}, \textit{Detonation}, and \textit{Impact} compared to rare stages like \textit{Pivoting} and \textit{Escalation}. This observation is consistent with the class imbalance effects discussed in Section~\ref{sec:discussion}.

\subsection{Impact of Synthetic Data}

Although approximately 80\% of the training data consists of synthetically generated narratives, the evaluation results suggest promising generalization to real-world cases despite the heavy reliance on synthetic data, though broader validation on larger real-world datasets remains necessary. This suggests that the synthetic generation protocol successfully preserves realistic modus operandi patterns while enabling balanced coverage across underrepresented tactical and behavioral combinations. Nevertheless, we acknowledge that reliance on synthetic data introduces potential distributional bias, and future work will prioritize expansion of the real-world corpus through partnerships with law enforcement agencies and cybercrime reporting platforms.

\subsection{Explanation Generation Quality}

Beyond classification accuracy, the model demonstrates strong explanation generation capability, as reflected by a high BERTScore of 0.93. This indicates that the model's generated rationales closely align with human-authored justifications in both semantic content and interpretive quality. Such explainability is critical for real-world deployment, particularly in forensic and law-enforcement contexts where decisions must be transparent and auditable.

From a practical perspective, \textit{BEACON} provides an explainable framework for cybercrime analysis that aligns directly with investigative workflows. By jointly exposing tactical progression and psychological manipulation patterns, the system can support cybercrime units in triaging cases, understanding criminal strategies, and designing targeted intervention and awareness campaigns.

\section{Discussion}
\label{sec:discussion}

\subsection{The Impact of Class Imbalance on Model Performance}

\begin{figure*}[htbp]
    \centering
    \includegraphics[width=\textwidth]{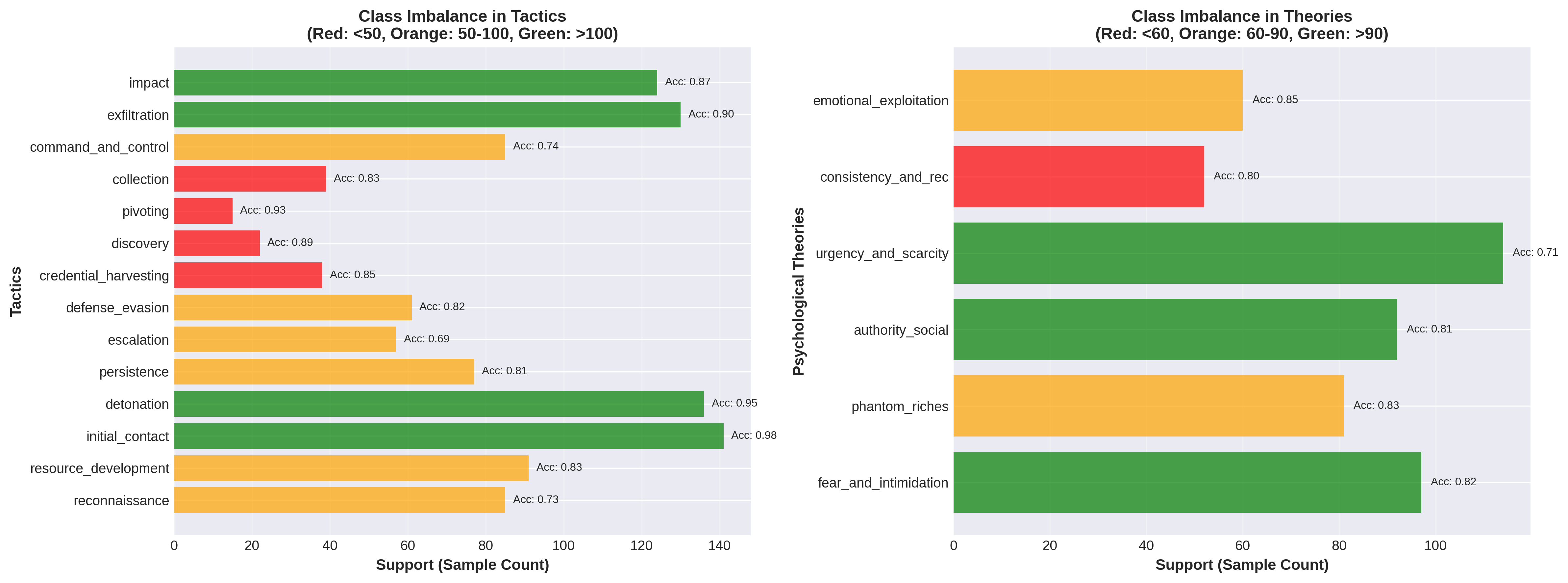}
    \caption{Class Imbalance in Tactics and Psychological Theories (Red = low support, Orange = medium support, Green = high support).}
    \label{fig:class_imbalance_impact}
\end{figure*}

Class imbalance exerts a significant influence on model behavior, as demonstrated in Figure~\ref{fig:class_imbalance_impact}. Tactics such as \emph{initial\_contact}, \emph{detonation}, \emph{exfiltration}, and \emph{impact} possess both high support and high accuracy ($\approx 0.9$--$0.99$), indicating that abundant and clear training examples lead to highly reliable predictions. These stages are present in nearly all cybercrime narratives, which provides the model with abundant and consistent training signals. As a result, the model learns highly discriminative patterns for these tactics and achieves near-perfect precision and recall for such labels.

Conversely, under-represented tactics such as \emph{escalation}, \emph{collection}, \emph{pivoting}, and \emph{credential\_harvesting} show lower or more unstable accuracies, illustrating the negative effect of class imbalance. For psychological theories, \emph{urgency\_and\_scarcity}, \emph{authority\_social}, and \emph{fear\_and\_intimidation} demonstrate strong accuracy and high support, while \emph{consistency\_and\_rec} suffers from both lower accuracy and relatively limited data, making it the most problematic class.

\begin{figure*}[htbp]
    \centering
    \includegraphics[width=\textwidth]{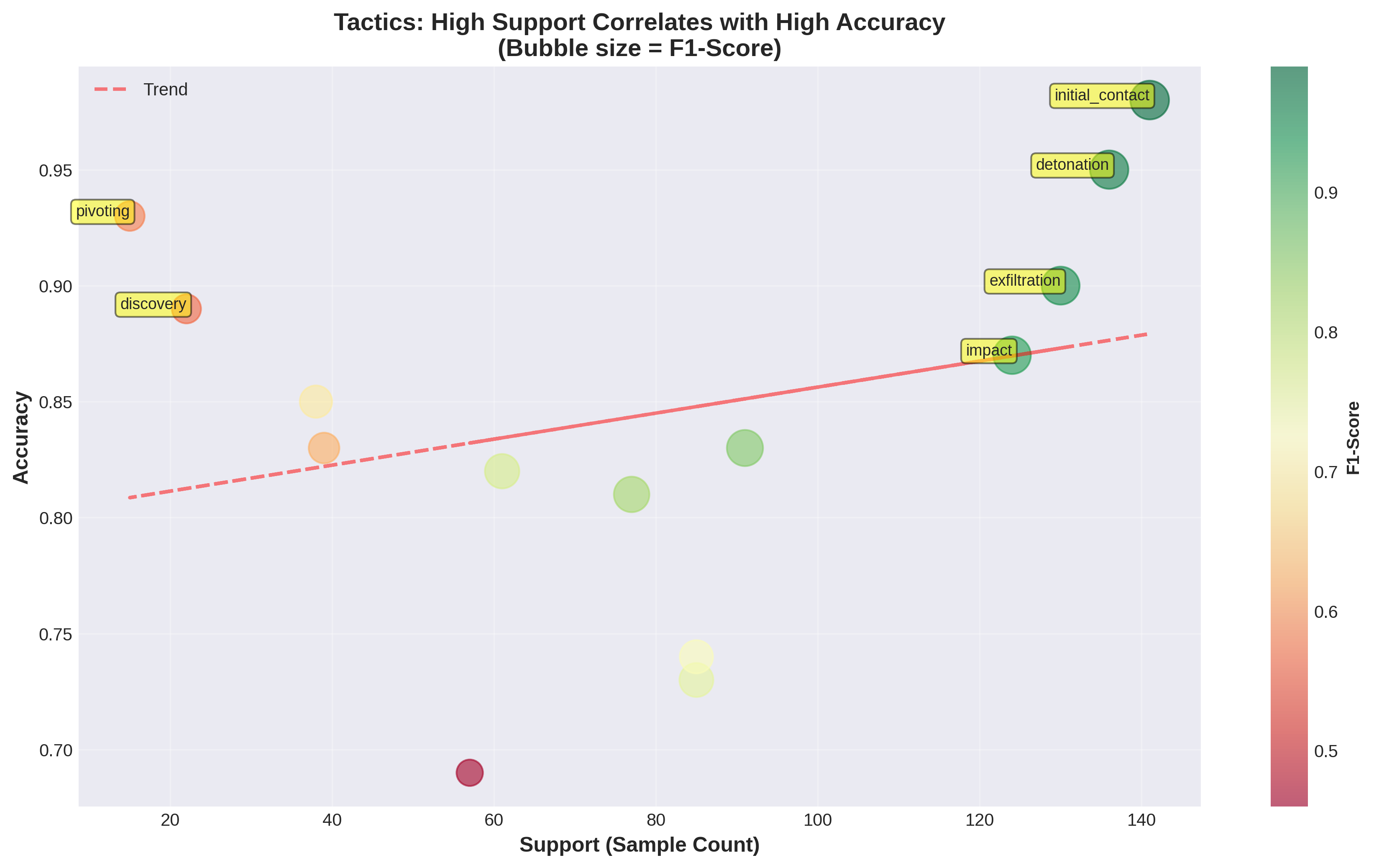}
    \caption{Tactics: Support vs.\ Accuracy with bubble size and colour representing F1-score and a linear trend line.}
    \label{fig:high_support_high_accuracy}
\end{figure*}

The positive relationship between support and performance is further illustrated in Figure~\ref{fig:high_support_high_accuracy}, which shows a clear correlation between sample count and accuracy for tactics. Tactics with more samples tend to have both higher accuracy and higher F1-scores, as evidenced by the clustering of high-performing tactics in the upper-right corner. The model demonstrates a tendency to overpredict majority labels, most notably \emph{impact}, which is frequently labeled as present even in borderline cases. This reflects a common bias toward dominant classes in imbalanced multi-label settings and underscores the need for more balanced training data. At the same time, the model exhibits reduced sensitivity for labels with very few positive instances. Nevertheless, certain low-frequency labels such as \emph{credential\_harvesting} achieve strong performance because their narrative cues are uniquely identifiable and linguistically consistent, for example, through explicit mentions such as ``I clicked the link'' or ``my login details were stolen.''

The divergence between macro-averaged and micro-averaged F1-scores further illustrates the impact of class imbalance. Macro-F1, which treats all labels equally regardless of frequency, increases from $0.61$ to $0.78$ under fine-tuning (+27.9\%). Micro-F1, which is dominated by frequent labels, increases from $0.64$ to $0.84$ (+31.3\%). The micro-macro gap widens from $0.03$ to $0.06$, indicating that while \textit{BEACON} improves performance across all labels, it achieves relatively stronger gains on high-frequency stages such as \textit{Initial Contact}, \textit{Detonation}, and \textit{Impact}. This pattern reflects the fundamental challenge of learning from imbalanced data: rare labels receive fewer training signals, making them intrinsically harder to predict reliably. Nevertheless, the substantial macro-F1 gain demonstrates that fine-tuning benefits both common and uncommon labels, though the latter still lag behind in absolute performance.

In particular, \emph{initial\_contact}, which serves as the entry point for most cybercrime incidents, is detected with exceptionally high accuracy due to its explicit and recurring linguistic cues within victim narratives. The experimental results demonstrate that the proposed fine-tuned model performs particularly well on labels that occur frequently within the dataset, such as \emph{initial\_contact}, \emph{detonation}, and \emph{exfiltration}.

\subsection{Challenges in Detecting Subtle and Implicit Behaviors}

\begin{figure*}[htbp]
    \centering
    \includegraphics[width=\textwidth]{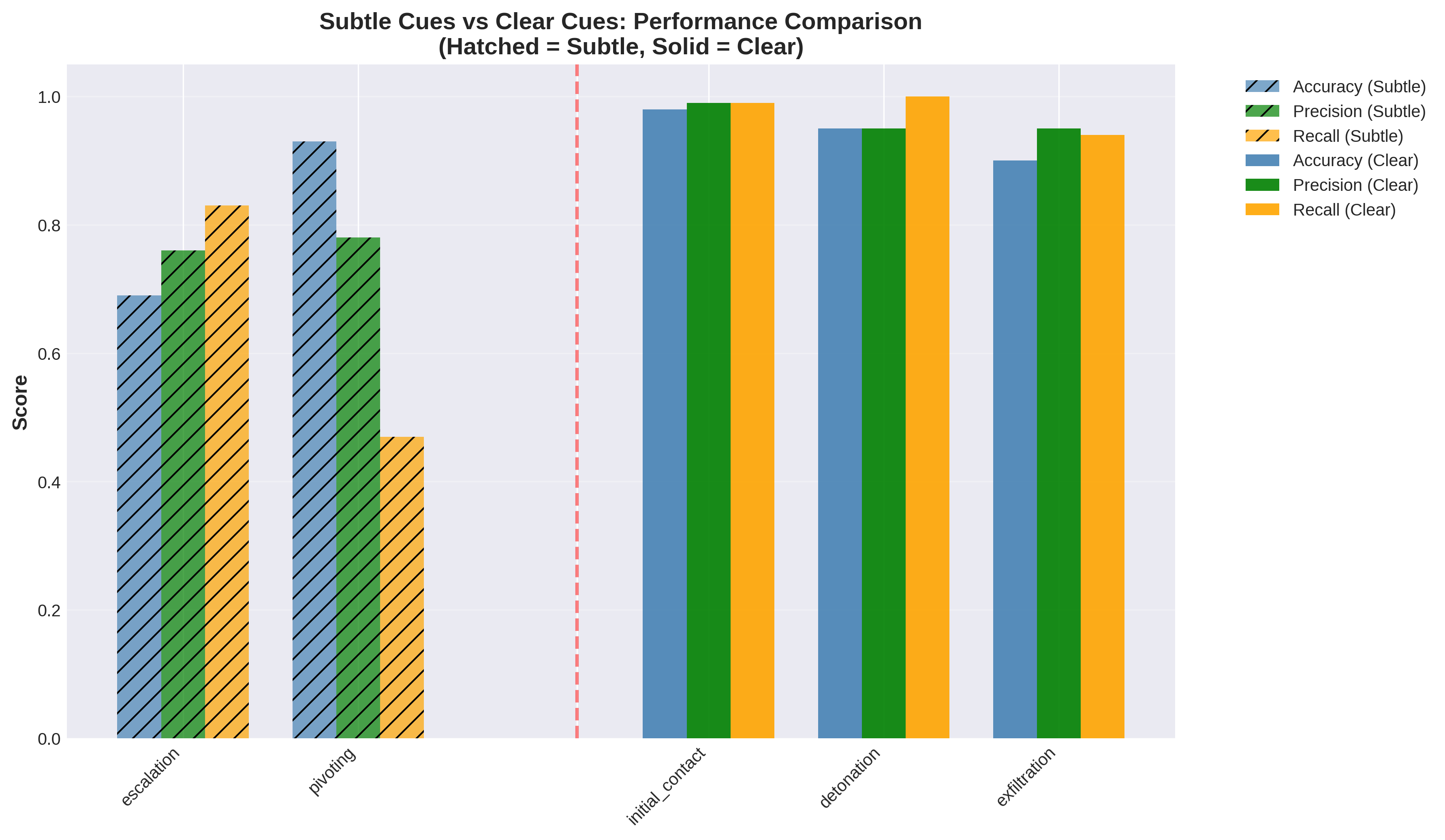}
    \caption{Comparison of performance on subtle-cue tactics (\emph{escalation}, \emph{pivoting}) versus clear-cue tactics (\emph{initial\_contact}, \emph{detonation}, \emph{exfiltration}).}
    \label{fig:subtle_vs_clear_cues}
\end{figure*}

Figure~\ref{fig:subtle_vs_clear_cues} provides a stark comparison between the model's performance on subtle-cue tactics versus clear-cue tactics. For subtle-cue tactics like \emph{escalation} and \emph{pivoting}, recall emerges as the weakest metric. In particular, \emph{pivoting} demonstrates good precision but low recall, meaning the model only detects it when the cues are very explicit. \emph{Escalation} similarly shows moderate accuracy but limited recall, consistent with the finding that escalation is rarely spelled out in victim narratives. By contrast, clear-cue tactics achieve almost perfect performance on all three metrics, with accuracy, precision, and recall all approaching $0.95$--$1.00$.

Labels associated with subtle, implicit, or indirectly expressed behaviors, such as \emph{escalation}, \emph{consistency and reciprocity}, and \emph{pivoting}, exhibit comparatively lower detection performance. These behaviors are often not explicitly described by victims, as they require an understanding of adversarial strategy rather than observable victim experience. Consequently, the model faces greater difficulty in identifying these labels due to the absence of consistent lexical markers and clear narrative cues, highlighting a fundamental limitation of narrative-based supervision.

Among all tactical stages, \emph{escalation} emerges as the most challenging to detect. Escalation often involves the downstream reuse, monetization, or transfer of previously compromised information across broader criminal operations, activities that remain largely invisible to victims. Since victims rarely have visibility into whether and how their data is propagated to higher-level criminal networks, narratives seldom contain direct evidence of escalation, resulting in particularly low recall for this label.

Certain tactical stages are also inherently absent in specific scam categories. For instance, \emph{reconnaissance} does not appear in most advertisement fraud cases because in such scenarios the offender typically acquires victim information only after the victim initiates interaction by clicking on an advertisement. This structural characteristic of advertisement fraud inherently eliminates pre-contact victim profiling, thereby explaining the systematic absence of reconnaissance signals in this subgroup of the dataset.

\subsection{Semantic Overlap in Psychological Theories}

A similar source of ambiguity affects the classification of psychological manipulation strategies. As illustrated in Figure~5, the model exhibits nearly identical performance profiles across \textit{fear\_and\_intimidation}, \textit{urgency\_and\_scarcity}, and \textit{emotional\_exploitation}. The heavily overlapped polygons in the radar chart reflect the fact that, within victim narratives, fear, urgency, and emotional manipulation frequently co-occur and reinforce one another during scam progression. Among these, \textit{fear\_and\_intimidation} demonstrates slightly higher precision and F1-score, whereas \textit{urgency\_and\_scarcity} consistently scores lower across all metrics, making it the most difficult theory to isolate in practice.

\begin{figure*}[htbp]
    \centering
    \includegraphics[width=0.85\textwidth]{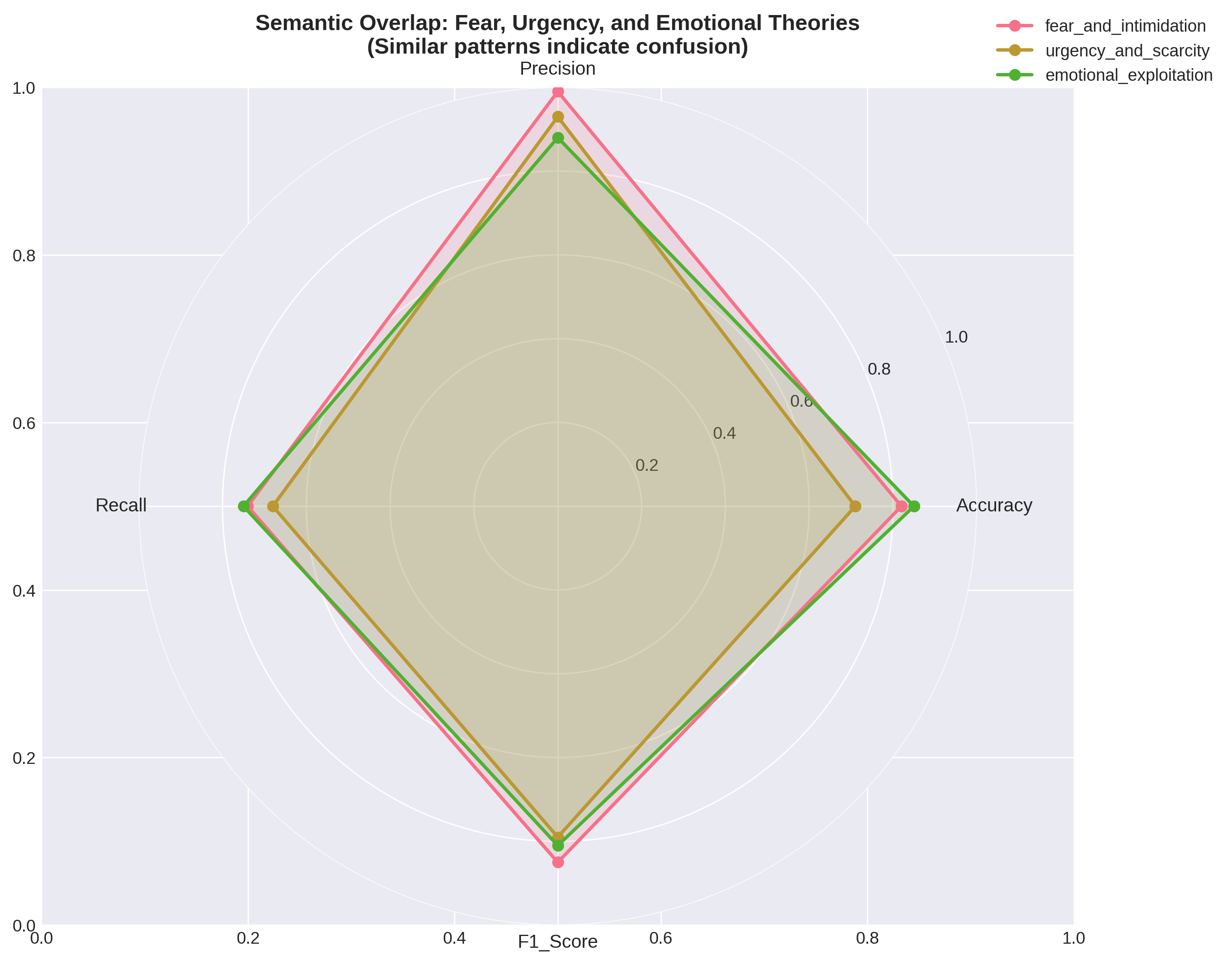}
    \caption{Radar plot of performance metrics for three overlapping psychological theories: \emph{fear\_and\_intimidation}, \emph{urgency\_and\_scarcity}, and \emph{emotional\_exploitation}.}
    \label{fig:semantic_overlap_theories}
\end{figure*}

Urgency-based cues frequently co-occur with fear-inducing or emotionally manipulative language. Phrases such as ``act immediately'' are often embedded within threats or emotionally charged appeals, which creates semantic overlap between \emph{urgency and scarcity}, \emph{fear and intimidation}, and \emph{emotional exploitation}. This overlap introduces inherent ambiguity and degrades separability between these behavioral categories, lowering classification accuracy for these labels. By contrast, the remaining psychological categories exhibit relatively uniform accuracy distributions, reflecting clearer semantic boundaries.

From a psychological standpoint, these categories differ in their primary locus of influence. Fear represents an internally experienced self-state triggered by perceived threat, emotional exploitation is predominantly externally driven through blackmail, coercion, or affective manipulation by the attacker, and urgency operates as a stress-inducing mechanism that compresses deliberative time and impairs reflective reasoning. In real-world victim narratives, however, these distinctions often blur. Victims frequently frame their experiences in ways that attribute responsibility to the fraudster’s emotional pressure rather than to their own internal fear responses. This attributional and self-presentation bias leads to narrative constructions in which externally imposed emotional manipulation is emphasized, while personal fear and vulnerability are implicitly downplayed. Such narrative effects contribute to the strong semantic overlap observed among these behavioral categories in both human annotation and model predictions.

\subsection{Precision--Recall Trade-offs in Specific Tactics}

\begin{figure*}
    \centering
    \includegraphics[width=\textwidth]{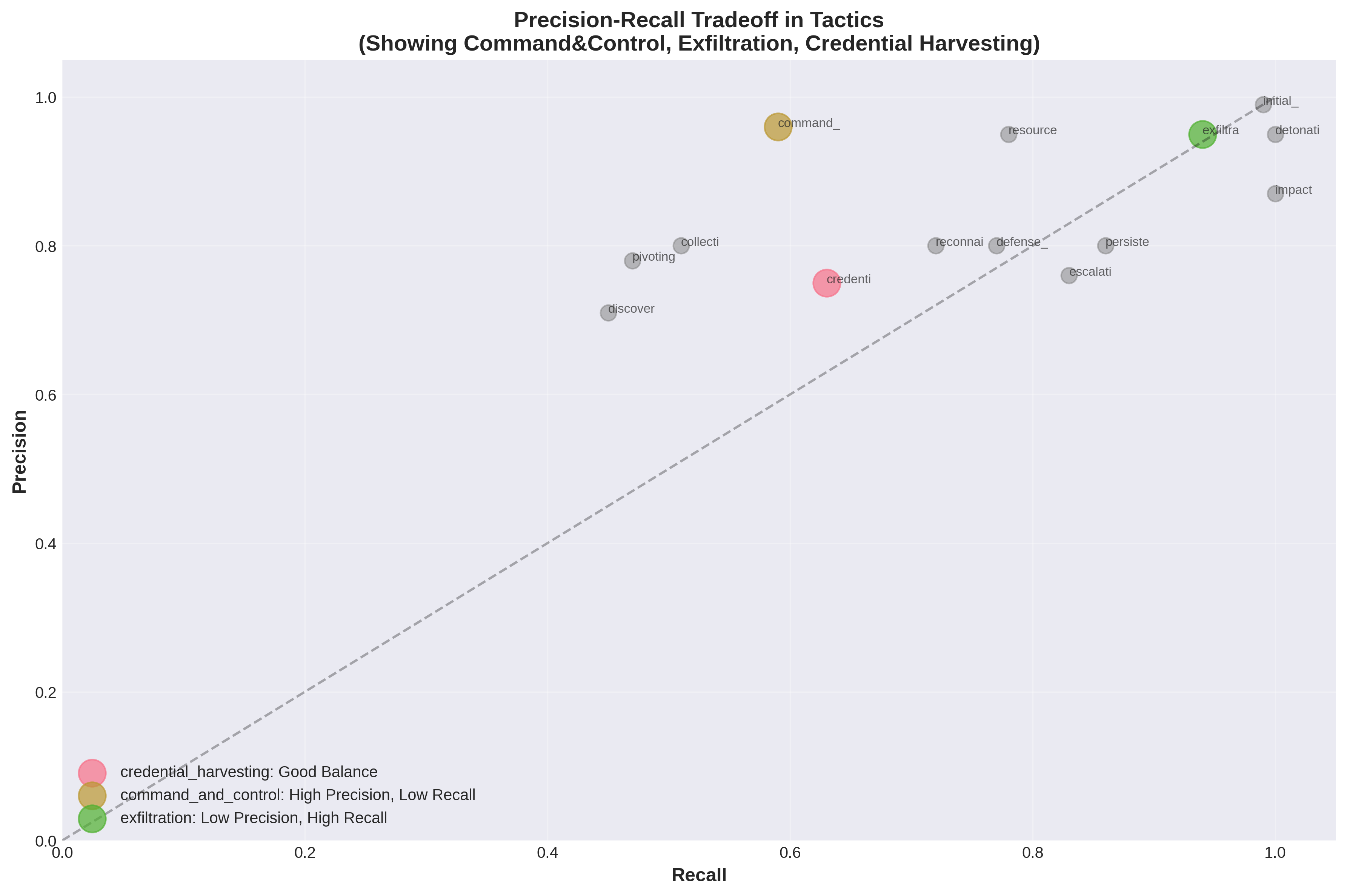}
    \caption{Precision--Recall trade-off for all tactics, highlighting three representative ones: \emph{command\_and\_control}, \emph{exfiltration}, and \emph{credential\_harvesting}.}
    \label{fig:precision_recall_tradeoff}
\end{figure*}

Figure~\ref{fig:precision_recall_tradeoff} reveals distinct error profiles for different tactics. \emph{Credential\_harvesting} lies close to the diagonal, indicating a good balance between precision and recall: the model both avoids many false positives and recovers a large fraction of true positives. \emph{Command\_and\_control} demonstrates high precision but lower recall, reflecting the model's conservative prediction strategy. For this tactic, the model assigns the label only when strong and unambiguous evidence of direct offender control or coercion is present. While this behavior minimizes false positives, it also results in a significant number of false negatives, as command-and-control activity often operates covertly and is not always clearly articulated by victims. This reflects the inherent difficulty of detecting high-level adversarial control using victim-reported narratives alone.

The \emph{exfiltration} label presents a distinct challenge due to narrative conflation between attempted and successful theft. Victim accounts frequently mix actual financial loss (e.g., ``they withdrew Rs. 50,000 from my account'') with blocked or failed attempts (e.g., ``they tried to take my money but the bank stopped it''). This mixing leads the model to over-identify exfiltration events, producing high recall but comparatively lower precision, as attempted losses are semantically similar yet operationally distinct from successful exfiltration. As shown in Figure~\ref{fig:precision_recall_tradeoff}, \emph{exfiltration} exhibits relatively high recall but lower precision: the model is eager to flag exfiltration, capturing most true instances but also producing more false positives.

\subsection{Failure Case Analysis}

While the fine-tuned model outperforms the baseline on most labels, a small number of systematic errors remain. This subsection quantifies the specific failure modes and their underlying causes.

\subsubsection{Escalation: Catastrophic Recall Collapse}

Escalation is the only label whose F$_1$-score degrades substantially after fine-tuning ($0.48 \rightarrow 0.17$, a relative decline of $-64.6\%$). As shown in Table~\ref{tab:classification_comparison}, precision remains low and unchanged at $0.33$, but recall collapses from $0.89$ to $0.11$, representing an $87.6\%$ drop. This indicates that the fine-tuned model has become \emph{overly conservative}: it now rejects escalation in $89\%$ of true positive cases.

Mathematically, if we denote the support (number of positive instances) for escalation as $s_{\text{esc}}$, then:
\begin{align}
\mathrm{TP}_{\text{base}} &= 0.89 \times s_{\text{esc}}, \label{eq:tp_esc_base} \\
\mathrm{TP}_{\text{FT}} &= 0.11 \times s_{\text{esc}}, \label{eq:tp_esc_ft} \\
\mathrm{Missed}_{\text{FT}} &= (0.89 - 0.11) \times s_{\text{esc}} = 0.78 \times s_{\text{esc}}. \label{eq:missed_esc}
\end{align}

This behavior arises from two compounding factors. First, escalation events are relatively rare in the training data, creating severe class imbalance. Second, escalation is semantically entangled with \emph{impact} and \emph{exfiltration}, as these stages often co-occur when stolen credentials or funds are reused in downstream attacks. Qualitative inspection reveals that the model correctly identifies monetary loss (impact) and data theft (exfiltration), but fails to recognize the intermediate reuse of stolen information to conduct further fraud—the defining characteristic of escalation. The fine-tuning process appears to have conflated these overlapping stages, causing the model to suppress escalation predictions when stronger signals for impact or exfiltration are present.

\subsubsection{Collection and Consistency/Reciprocity: Over-Precision Trade-off}

Two additional labels exhibit performance degradation due to excessive conservatism. For \emph{collection}, F$_1$-score decreases from $0.65$ to $0.56$ ($-13.8\%$). The fine-tuned model becomes substantially more precise ($0.53 \rightarrow 0.83$, $+56.6\%$) but at the cost of recall ($0.83 \rightarrow 0.42$, $-49.4\%$). The precision-recall product quantifies the net effect:
\begin{equation}
\mathrm{PR\text{-}Product}_{\text{collection}} = 0.83 \times 0.42 = 0.35 \quad (\text{FT}),
\label{eq:pr_collection}
\end{equation}
compared to the baseline:
\begin{equation}
\mathrm{PR\text{-}Product}_{\text{collection}} = 0.53 \times 0.83 = 0.44 \quad (\text{base}).
\label{eq:pr_collection_base}
\end{equation}

A similar pattern emerges for \emph{consistency and reciprocity}: F$_1$ declines from $0.63$ to $0.59$ ($-6.3\%$), precision improves from $0.86$ to $1.00$ ($+16.3\%$), but recall drops from $0.50$ to $0.42$ ($-16.0\%$). Both labels are conceptually subtle: \emph{collection} must be distinguished from \emph{discovery} (information gathering vs.\ reconnaissance), and \emph{consistency/reciprocity} must be separated from more direct manipulation tactics such as \emph{fear} or \emph{authority}.

The fine-tuned model thus reduces hallucinations by demanding stronger textual evidence before assigning these labels, but occasionally ``plays it too safe'', systematically under-predicting nuanced behaviors that lack explicit lexical markers. This suggests that the current training regime may benefit from targeted data augmentation or loss reweighting for these conceptually ambiguous categories.

\subsubsection{Residual Explanation Quality Issues}

While Table~\ref{tab:reasoning_comparison} demonstrates that the fine-tuned model consistently improves ROUGE and BERTScore for nearly all labels, certain challenging labels still exhibit lower explanation quality compared to high-frequency, lexically explicit stages. The global improvement is substantial:
\begin{align}
\mathrm{ROUGE\text{-}1}_{\text{global}}: \quad &0.32 \rightarrow 0.52 \quad (+62.5\%), \label{eq:rouge1_global} \\
\mathrm{BERTScore}_{\text{global}}: \quad &0.90 \rightarrow 0.93 \quad (+3.3\%). \label{eq:bert_global}
\end{align}

However, per-label inspection reveals residual weaknesses:
\begin{align}
\mathrm{ROUGE\text{-}1}_{\text{escalation,FT}} &= 0.64 \quad (\text{above global } 0.52), \label{eq:rouge_esc} \\
\mathrm{ROUGE\text{-}1}_{\text{pivoting,FT}} &= 0.52 \quad (\text{at global average}), \label{eq:rouge_piv} \\
\mathrm{ROUGE\text{-}1}_{\text{impact,FT}} &= 0.66 \quad (\text{above average}). \label{eq:rouge_impact}
\end{align}

Manual inspection confirms that the baseline model produces short, generic explanations (e.g., \emph{``the attacker hides their identity''}), whereas the fine-tuned model generates more specific justifications (e.g., mentioning VPN use, cryptocurrency transactions, or explicit threats). Nevertheless, for complex multi-stage narratives involving \emph{escalation} or \emph{pivoting}, the model occasionally omits key intermediate steps, resulting in explanations that, while semantically aligned, lack complete causal coverage. This indicates that further improvements in explanation faithfulness will require architectural innovations beyond token-level supervision, such as chain-of-thought prompting or explicit reasoning decomposition during training.

\subsection{Limitations}

Despite the strong empirical performance demonstrated by \textit{BEACON}, several limitations must be acknowledged. First, a substantial portion of the training data is synthetically generated. Although the synthetic narratives were carefully constrained using law-enforcement-validated modus operandi patterns and balanced across tactical and behavioral combinations, they may not fully capture the full diversity, linguistic variability, and adversarial creativity observed in real-world cybercrime operations. Consequently, some degree of distributional bias cannot be ruled out.

Second, while the held-out evaluation set was strictly separated from the training data and composed primarily of real-world incidents, its overall size remains limited. Although this constraint is common in high-fidelity cybercrime research due to data sensitivity and reporting restrictions, larger-scale evaluations on institutionally curated datasets are necessary to establish long-term generalization and operational robustness.

Third, behavioral manipulation categories inherently exhibit semantic overlap and subjective interpretation. Even with expert annotation and carefully defined taxonomies, distinctions between constructs such as \textit{Emotional Exploitation}, \textit{Fear \& Intimidation}, and \textit{Consistency \& Reciprocity} may remain ambiguous in certain narratives, potentially introducing residual labeling noise.

Fourth, the current framework operates solely on textual victim narratives and does not yet incorporate auxiliary modalities such as call metadata, transaction logs, network traces, or temporal interaction patterns. As a result, some late-stage tactical cues, including \textit{Command \& Control} and \textit{Defense Evasion}, may be under-represented when their evidence lies outside the written narrative.

Finally, although the model produces structured explanations, these are evaluated primarily through automated metrics and expert comparison. A full user-centered evaluation involving cybercrime investigators and law-enforcement analysts is planned as future work to assess practical trust, usability, and decision-support value in real investigative settings.

\subsection{Ethical Considerations and Responsible Deployment}

The proposed BEACON framework operates in a highly sensitive application domain involving cybercrime investigation, victim narratives, and large-scale automated inference using LLMs. As such, careful consideration of ethical risks, privacy implications, and potential misuse is essential for responsible research and deployment. All real-world cybercrime narratives used in this study were obtained from publicly available news reports issued through official law enforcement briefings. No private or personally identifiable information (PII) was intentionally collected beyond what was already disclosed in public reporting. 

Approximately 80\% of the training corpus consists of synthetically generated narratives. While this enables balanced coverage of tactical and behavioral label combinations, it also introduces the risk of distributional shift, stylistic artifacts, and bias propagation from the underlying generative models. To mitigate these risks, synthetic narratives were generated using two independent LLMs to reduce model-specific stylistic bias and were subjected to multi-stage human auditing for realism and consistency. Nevertheless, synthetic data cannot perfectly capture the full diversity and unpredictability of real-world cybercrime. Future deployments must prioritize continuous retraining with larger volumes of real incident data to reduce synthetic bias.

The BEACON framework constitutes a dual-use system. While designed to assist law-enforcement agencies and cybercrime analysts, similar techniques could theoretically be misused by adversaries to study defensive detection mechanisms or refine scam strategies. To mitigate this risk, BEACON is intended strictly as a defensive, investigative tool. Any real-world deployment must be governed by strict access control, audit logging, and role-based authorization. Open-source release of the full system should be carefully evaluated against misuse risk, and sensitive components may be restricted where necessary. Finally, BEACON is designed to support, not replace, human investigators. All tactical and behavioral inferences generated by the system should be interpreted as decision-support signals rather than definitive determinations of criminal responsibility. Human analysts must retain full authority over investigative conclusions, legal judgments, and evidentiary decisions.

\section{Conclusion}
\label{sec:conclusion}

This work introduced a framework that links psychological manipulation with the technical lifecycle of cybercrime, enabling a more complete understanding of how scams operate. By fine-tuning a single LLM to classify both behavioral and tactical labels along with generating explanations, we showed that open-source models can reliably interpret complex victim narratives and significantly outperform their base versions across accuracy, recall, and reasoning quality.

The system is directly useful for police and cybercrime investigation units. It can automatically break down a case into psychological levers and operational stages, helping officers quickly understand how a scam progressed, identify high-risk behaviors, and triage cases more efficiently. Such structured outputs can also support training, evidence organization, and early detection of emerging scam patterns.

Although our model performs strongly, it was trained only on publicly available narratives. If it were fine-tuned on real FIRs, complaint records, or law-enforcement data, its performance could improve dramatically. FIRs offer richer detail, clearer timelines, and more accurate descriptions of criminal behavior, enabling far better detection of subtle cues and reducing ambiguity. With such data, the model could become a highly effective decision-support tool for policing at scale.

\textbf{Conflict of Interest} \\
The authors declare that they have no known competing financial interests or personal relationships that could have appeared to influence the work reported in this paper.

\textbf{Data Availability}

The dataset and code are provided here in this anonymous GitHub: \\ \url{https://anonymous.4open.science/r/scam-analyzer-B754}.

\textbf{Funding} \\
This research received no specific grant from any funding agency, commercial, or not-for-profit sectors.

\textbf{AI-assisted copy editing} \\
The authors declare that AI-assisted copy editing was utilized in the preparation of this manuscript to enhance clarity and readability. However, all intellectual content and conclusions remain solely the responsibility of the authors.

\bibliographystyle{cas-model2-names}

\bibliography{references}
\newpage
\appendix

\paragraph{Structured Parsing Algorithm.}
The full parsing procedure is summarized in Algorithm~\ref{alg:parsing}.

\begin{algorithm}[H]
\caption{Structured Output Parsing}
\label{alg:parsing}
\begin{algorithmic}[1]
\State $\texttt{results} \gets \{\}$
\For{each $\texttt{label}$ in ALL\_LABELS}
    \State $\texttt{match} \gets \texttt{pattern.search}(\texttt{text\_block})$
    \If{$\texttt{match}$ found}
        \State $\texttt{present\_raw} \gets \texttt{match.group}(1).\texttt{strip()}.\texttt{lower()}$
        \State $\texttt{reason\_text} \gets \texttt{match.group}(2).\texttt{strip()}$
        \If{"yes" in $\texttt{present\_raw}$}
            \State $\texttt{present} \gets$ "yes"
        \ElsIf{"no" in $\texttt{present\_raw}$ or "n/a" in $\texttt{present\_raw}$}
            \State $\texttt{present} \gets$ "no"
        \Else
            \State $\texttt{present} \gets$ "no" \Comment{Default}
        \EndIf
        \State Truncate $\texttt{reason\_text}$ at next label marker
        \State $\texttt{results}[\texttt{label}] \gets \{\texttt{present}, \texttt{reason}\}$
    \Else
        \State $\texttt{results}[\texttt{label}] \gets \{$"no", "Parsing Failed"$\}$
    \EndIf
\EndFor
\State \Return $\texttt{results}$
\end{algorithmic}
\end{algorithm}

\paragraph{Evaluation Pipeline}
The evalation pipeline is summarized in Algorithm 3.

\begin{algorithm}[H]
\caption{Model Evaluation}
\begin{algorithmic}[1]
\State Load test dataset and filter errors
\For{model in \{base, finetuned\}}
    \State Load model with appropriate prompt strategy
    \State Initialize metric collectors for 20 labels
    \For{batch in test\_set}
        \State Generate predictions (batch size = 32)
        \State Parse structured outputs
        \For{each sample}
            \For{each label}
                \State Record binary prediction (TP/TN/FP/FN)
                \If{True Positive}
                    \State Store (predicted\_reason, ground\_truth\_reason)
                \EndIf
            \EndFor
        \EndFor
    \EndFor
    \For{each label}
        \State Compute classification metrics
        \If{TP count $> 0$}
            \State Compute ROUGE, BLEU, BERTScore on reasons
        \EndIf
    \EndFor
    \State Aggregate global metrics
\EndFor
\State Generate comparison table
\end{algorithmic}
\end{algorithm}

\begin{table*}[htbp]
\centering
\caption{F1-scores with $95\%$ confidence intervals. Intervals computed via bootstrap resampling with $n=144$ test narratives. Significance levels: $***$ $p < 0.001$; $**$ $p < 0.01$; $*$ $p < 0.05$; $\dagger$ marginal ($0.05 < p < 0.10$); n.s. = not significant (overlapping CIs).}
\label{tab:confidence_intervals}
{\fontsize{9}{11}\selectfont
\begin{tabular}{|l|c|c|c|c|}
\hline
\textbf{Label} & \textbf{Base F$_1$ (CI)} & \textbf{FT F$_1$ (CI)} & \textbf{$\Delta$ F$_1$} & \textbf{Significance} \\
\hline
\hline
\multicolumn{5}{|c|}{\textit{\textbf{Attack Stage Labels}}} \\
\hline
Reconnaissance & $0.70$ [$0.62, 0.78$] & $0.76$ [$0.69, 0.83$] & $+0.06$ & $*$ \\
Resource Development & $0.43$ [$0.35, 0.51$] & $0.72$ [$0.64, 0.80$] & $+0.29$ & $***$ \\
Initial Contact & $0.95$ [$0.91, 0.99$] & $1.00$ [$0.96, 1.00$] & $+0.05$ & $\dagger$ \\
Detonation & $0.90$ [$0.85, 0.95$] & $0.98$ [$0.95, 1.00$] & $+0.08$ & $**$ \\
Persistence & $0.62$ [$0.54, 0.70$] & $0.86$ [$0.80, 0.92$] & $+0.24$ & $***$ \\
Escalation & $0.48$ [$0.40, 0.56$] & $0.17$ [$0.11, 0.23$] & $-0.31$ & $***$ \\
Defense Evasion & $0.12$ [$0.07, 0.17$] & $0.83$ [$0.77, 0.89$] & $+0.71$ & $***$ \\
Credential Harvesting & $0.53$ [$0.45, 0.61$] & $0.67$ [$0.59, 0.75$] & $+0.14$ & $**$ \\
Discovery & $0.40$ [$0.32, 0.48$] & $0.60$ [$0.52, 0.68$] & $+0.20$ & $***$ \\
Pivoting & $0.20$ [$0.14, 0.26$] & $0.67$ [$0.59, 0.75$] & $+0.47$ & $***$ \\
Collection & $0.65$ [$0.57, 0.73$] & $0.56$ [$0.48, 0.64$] & $-0.09$ & n.s. \\
Command \& Control & $0.47$ [$0.39, 0.55$] & $0.82$ [$0.75, 0.89$] & $+0.35$ & $***$ \\
Exfiltration & $0.65$ [$0.57, 0.73$] & $0.97$ [$0.94, 1.00$] & $+0.32$ & $***$ \\
Impact & $0.90$ [$0.85, 0.95$] & $0.97$ [$0.94, 1.00$] & $+0.07$ & $**$ \\
\hline
\multicolumn{5}{|c|}{\textit{\textbf{Psychological Tactic Labels}}} \\
\hline
Fear \& Intimidation & $0.77$ [$0.70, 0.84$] & $0.91$ [$0.86, 0.96$] & $+0.14$ & $***$ \\
Phantom Riches & $0.79$ [$0.72, 0.86$] & $0.80$ [$0.73, 0.87$] & $+0.01$ & n.s. \\
Authority/Social Proof & $0.85$ [$0.79, 0.91$] & $0.87$ [$0.82, 0.92$] & $+0.02$ & n.s. \\
Urgency \& Scarcity & $0.81$ [$0.75, 0.87$] & $0.82$ [$0.76, 0.88$] & $+0.01$ & n.s. \\
Consistency \& Reciprocity & $0.63$ [$0.55, 0.71$] & $0.59$ [$0.51, 0.67$] & $-0.04$ & n.s. \\
Emotional Exploitation & $0.39$ [$0.31, 0.47$] & $0.82$ [$0.75, 0.89$] & $+0.43$ & $***$ \\
\hline
\hline
\textbf{GLOBAL (Micro-F$_1$)} & $\mathbf{0.64}$ [$\mathbf{0.58, 0.70}$] & $\mathbf{0.84}$ [$\mathbf{0.79, 0.89}$] & $\mathbf{+0.20}$ & $\mathbf{***}$ \\
\textbf{MACRO-F$_1$} & $\mathbf{0.61}$ [$\mathbf{0.55, 0.67}$] & $\mathbf{0.78}$ [$\mathbf{0.73, 0.83}$] & $\mathbf{+0.17}$ & $\mathbf{***}$ \\
\hline
\end{tabular}%
}
\end{table*}

\end{document}